\documentclass[acmsmall]{acmart}

\usepackage[normalem]{ulem}

\usepackage[utf8]{inputenc}
\usepackage[T1]{fontenc}

\usepackage{appendix}
\usepackage{pdfpages}
 
\usepackage{booktabs}
\usepackage{tabularx}
\usepackage{makecell}

\usepackage{graphicx}
\graphicspath{ {figures/} }
\usepackage{tikz}
\usetikzlibrary{positioning}
\newcommand*\RedSolid[1][1ex]{\tikz{\draw[->, red,solid,line width=1.5pt](0,0) -- (10mm,0);}}
\newcommand*\RedDashed[1][1ex]{\tikz{\draw[->, red,dashed,line width=1.5pt](0,0) -- (10mm,0);}}
\newcommand*\BlackSolid[1][1ex]{\tikz{\draw[->, black,solid,line width=1.5pt, double distance=.01pt](0,0) -- (10mm,0);}}
\newcommand*\BlackDashed[1][1ex]{\tikz{\draw[->, black,dashed,line width=1.5pt](0,0) -- (10mm,0);}}
\newcommand*\blueSolid[1][1ex]{\tikz{\draw[->, blue,solid,line width=1.5pt](0,0) -- (10mm,0);}}
\newcommand*\blueDouble[1][1ex]{\tikz{\draw[->, blue,solid,line width=1.5pt, double distance= .01pt](0,0) -- (10mm,0);}}
\newcommand*\YellowSolid[1][1ex]{\tikz{\draw[->, yellow,solid,line width=1.5pt, double distance=0.01](0,0) -- (10mm,0);}}
\newcommand*\YellowDashed[1][1ex]{\tikz{\draw[->, yellow,dotted,line width=3pt](0,0) -- (10mm,0);}}

\setcopyright{rightsretained}
\acmJournal{PACMHCI}
\acmYear{2024}
\acmVolume{8}
\acmNumber{CSCW2}
\acmArticle{368}
\acmMonth{11}
\acmDOI{10.1145/3686907}

\makeatletter
\gdef\@copyrightpermission{
  \begin{minipage}{0.2\columnwidth}
   \href{https://creativecommons.org/licenses/by/4.0/}{\includegraphics[width=0.90\textwidth]{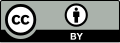}}
  \end{minipage}\hfill
  \begin{minipage}{0.8\columnwidth}
   \href{https://creativecommons.org/licenses/by/4.0/}{This work is licensed under a Creative Commons Attribution International 4.0 License.}
  \end{minipage}
  \vspace{5pt}
}
\makeatother

\begin{document}

\title{Security and Privacy Perspectives of People Living in Shared Home Environments}

\author{Nandita~Pattnaik}
\email{np407@kent.ac.uk}
\orcid{0000-0003-1272-077X}
\author{Shujun Li}
\email{S.J.Li@kent.ac.uk}
\orcid{0000-0001-5628-7328}
\author{Jason R.C.~Nurse}
\email{J.R.C.Nurse@kent.ac.uk}
\orcid{0000-0003-4118-1680}
\affiliation{%
  \department{Institute of Cyber Security for Society (iCSS) \& School of Computing}
  \institution{University of Kent}
  \city{Canterbury}
  \country{UK}
  \postcode{CT2 7NP}
  }

\begin{abstract}
Security and privacy (S\&P) perspectives of people in a multi-user home are a growing area of research, with many researchers reflecting on the complicated power imbalance and challenging access control issues of the devices involved. However, these studies primarily focused on the multi-user scenarios in traditional family home settings, leaving other types of multi-user home environments, such as homes shared by co-habitants without a familial relationship, under-studied. This paper closes this research gap via quantitative and qualitative analysis of results from an online survey and qualitative content analysis of sampled online posts on Reddit. The study explores the complex roles of shared home users, which depend on various factors unique to the shared home environment, e.g., who owns what home devices, how home devices are used by multiple users, and more complicated relationships between the landlord and people in the shared home and among co-habitants. Half (50.7\%) of our survey participants thought that devices in a shared home are less secure than in a traditional family home. This perception was found statistically significantly associated with factors such as the fear of devices being tampered with in their absence and (lack of) trust in other co-habitants and their visitors. We observed cyber-physical threats being a prominent topic discussed in Reddit posts. Our study revealed new user types and user relationships in a multi-user environment such as \textit{ExternalPrimary-InternalPrimary} while analysing the landlord and shared home resident relationship with regard to shared home device use. Based on the results of the online survey and the Reddit data, we propose a threat actor model for shared home environments, which has a focus on possible malicious behaviours of current and past co-habitants of a shared home, as a special type of \textit{insider threat} in a home environment. We also recommend further research to understand the complex roles co-habitants can play in navigating and adapting to a shared home environment's security and privacy landscape.
\end{abstract}



\maketitle

\section{Introduction}
\label{Sec:Intro}

``\textit{I have a smart house set up, Hue and Cync lights, Ring doorbell, myQ garage, Nest Thermostat and a Nest Hub Max in my room. I had a roommate who moved out on bad terms and still has legal rights to access our shared property when I am not home. How can I control his access so that he can only have access to my smart light and nothing else when he enters my home in my absence? I don't want him spying on my stuff.}''\footnote{The message is rephrased to be different from the original message while keeping its essence. This is necessary to avoid re-identification of the author of the post via a simple online search, as part of our research ethics application approved by our university.} The scenario paints a typical privacy issue faced by a shared house resident posted on the social media platform Reddit. It depicts a picture of how the traditional notion of household life is changing fundamentally from a closed, physical and private space to a hybrid half-public and smart living place characterised by `digitalization, connectedness, smartization and automation'~\cite{zhao2020unraveling}. This increasing digitisation is powered by the growing adoption of multiple smart and other traditional devices by household members~\cite{Statista2021AvgNoOfDevice}. For instance, a 2020 report from the British telecommunications operator BT~\cite{BT2020ConnectedDevice} stated that an average UK household possesses 28 connected devices on average. This burgeoning use of computing devices by home users brings about various security \& privacy (S\&P) related challenges, which have been widely discussed in the research literature~\cite{zeng2017end, zeng2019HomeUser, wickramasinghe2019survey, bernd2020bystanders, zimmermann2019assessing, lau2018alexa, Ahmad_2020_TangibleMuser, malkin2019privacy, marky2022PrivacyTea}. One of the primary factors contributing to the S\&P challenges of this growing conglomeration of devices is the number of users that own/use/share those devices. According to a piece of research by the Pew Research Center in 2020~\cite{PewResearchCenter2020home}, the average household size was 4.9 around the world, although this varies sharply regionally. The issues and challenges of a multi-user modern smart home are a subject of active research. Studies have focused on different topics while exploring this subject, which includes topics on different types of multi-user relationships while using those devices~\cite{zeng2019HomeUser, geeng2019Multi_User, Markey_2020_Idont_MultiUser, Cobb_2021_MultiUser, Ahmad_2020_TangibleMuser}, on the challenges and concerns they face~\cite{bernd2020bystanders, Apthorpe2022MU, Bernd2022MU, Kanchi2021MU, geeng2019Multi_User}, on user awareness and perceived norms~\cite{yao2019privacy, Ahmad_2020_TangibleMuser, Markey_2020_YouJust_MultiUser} and on mitigation techniques~\cite{huang_2020_amazon, lau2018alexa, yao2019privacy, Watson2020MU, Cobb2021MU, Ahmad_2020_TangibleMuser}.

However, the \textbf{context of most of these discussions principally referred to traditional family home settings}, where household members mainly include family members, visitors such as friends and guests, and people providing on-site household services such as nannies. This leaves a whole range of different non-family contexts that have not been covered by past research. For example, \textit{hybrid and extended homes}~\cite{zhao2020unraveling} where the definition of the home is more detached from a physical geo-location, \textit{pseudo-homes}~\cite{pattnaik2023survey} covering people who split their stay between two or more regular sites (e.g., students who live in their parent's house during term breaks and at university accommodation during term time), and \textit{shared homes}~\cite{Maalsen2020SharedHousing} where not all co-habitants are related by a familial bond, has not been addressed in past research as separate non-family contexts. These non-family home settings could lead to different types of S\&P concerns of users in comparison with family settings. In most traditional family-home settings, we have observed that the primary users have more overall control of the home devices~\cite{zeng2017end, geeng2019Multi_User} and, therefore, are the likely vulnerable points in home security. In a non-family home, each participating user is considered a primary user, and consequently, each of those primary users could expose the S\&P vulnerability of the shared home. Moreover, with the presence of multiple primary users, the negotiation of S\&P behaviours becomes trickier in a shared home environment. Additionally, each co-habitant might possess their own set of devices, such as a smart camera, adding not only to the number of devices and consequent S\&P vulnerabilities of a multi-device context but also creating potential new S\&P issues for other more passive users of such home devices. The S\&P perspectives of home users in these settings, therefore, should be researched separately from the users who live in a family setting. 

Past research~\cite{wilkinson2019stranger} has commented on the passing about concerns of shared home residents on physical safety and reluctance to stay in shared housing due to privacy concerns. However, there is a lack of research to understand S\&P-related multi-user issues and challenges that users of such houses might encounter as these types of households reflect a different way of living in contrast to the normativity of the nuclear family~\cite{maalsen2023SH}. For example, the multi-user relationships concerning user roles and power imbalances, i.e., between the primary and secondary users as defined in a traditional family unit~\cite{zeng2019HomeUser, geeng2019Multi_User, Markey_2020_Idont_MultiUser, Cobb_2021_MultiUser, Ahmad_2020_TangibleMuser} might not be applicable to a shared home owing to the existence of multiple primary users for common shared devices such as routers, smart locks or smart lights. The S\&P issues and concerns of the users in a non-family setting are affected by the fact that each co-habitant in a shared home setting often presents a different circle of outsiders; hence, the possible S\&P risks and threat actors in a non-family shared home setting are more likely higher than those in a family home setting. Furthermore, past research~\cite{nethercote2019caring} has highlighted how shared home living requires more frequent and dynamic negotiations about privacy boundaries in terms of domestic space or practices, unlike users in a family home setting, which raises the question of how these negotiations could influence S\&P behaviours of users in such non-family shared home settings. Approaches to S\&P vary depending on the context where it is experienced~\cite{wu2022sok}. Therefore, a thorough understanding of S\&P perspectives in various contexts is important to derive a comprehensive S\&P-related understanding regarding the users in multi-user homes.

This study, therefore, specifically focuses on the non-family S\&P perspectives of shared home users, as this type of shared living are becoming more common in some parts of the modern society with more people living in for a longer period and across a widening age demographics~\cite{Maalsen2020SharedHousing}. Such shared homes offer social, economic, and environmental benefits to their co-habitants and society as a whole. Similar evidence has been reported by others, e.g., a 2023 research paper~\citet{ronald2023institutionalization} reported that housing options such as shared accommodation and commercial co-living are fast becoming popular and increasingly institutionalised in European cities~\cite{ronald2023institutionalization} and the number of HMOs (houses in multiple occupations) increased by 22\% in 2021 from 2020 in London~\cite{London2022HMO}. This trend of increasing shared housing is seen globally in many countries and regions such as the UK~\cite{wilkinson2014compare}, Europe~\cite{uyttebrouck2020shared}, Japan~\cite{druta2021SH}, and New Zealand~\cite{clark2019rosters}, especially in large cities. One interesting aspect of this type of household is the diversity of the household members. According to recent data from a popular room share website as reported in a 2022 article on the BBC Magazine~\cite{BBC2022SharedHome}, the number of people who preferred to live in shared homes increased across different age bands, e.g., as much as 239\% for 55-64-year-olds, 114\% for 45-54-year-olds, 161\% for 65-year-olds, and 106\% for 35-44-year-olds. S\&P perspectives of people living in these homes, therefore, could help improve our understanding on cyber security and privacy aspects in such special types of home environments with multiple users from different backgrounds who often share many different devices. We therefore defined the following two research questions for our work.

\begin{itemize}
\item \textbf{RQ1}: What are the S\&P awareness and behaviours of people living in a shared home?

\item \textbf{RQ2}: What S\&P concerns are experienced by shared home co-habitants, and which threat actors are the likely sources of these concerns?
\end{itemize}

The above research questions were studied using a mixed research method: an online survey and analysis of data collected from a popular online forum (Reddit). Focusing on this specific, our main \textbf{findings and contributions} can be summarised as follows.
\begin{enumerate}
\item The results of our online survey showed that most of the survey participants (85\%) in a shared home reported to trust their co-habitants, but almost 50\% of them felt that a shared home setting is less secure compared to a traditional one about a single family.

\item We found that existing S\&P-related relationships between co-habitants in a shared home could be very complicated depending on the devices used.

\item The study also demonstrated the unique role of the landlord in most scenarios, where they are a primary user of some home devices from outside of the shared home. For instance, most people in a shared home equipped with shared smart devices owned by their landlords were never provided with any details of the nature of data collection, storage, or sharing and never signed any agreement in relation to this.

\item We found cyber-physical harm to be a frequent theme amongst those residents who were concerned about their co-habitants in many shared homes. These inhabitants were afraid of physical harm from their cohabitants, often exacerbated by the use of in-home smart devices, and were looking for a smart solution to the problem.

\item The study highlighted different S\&P perspectives of multiple device users in a shared home. For instance, it revealed the co-existsnce of different types of primary users of some devices, the balance of power between the co-habitants that exist in such a context, and the operational/usage problems that might arise due to the clash of multiple primary users.

\item We proposed a customised threat actor model highlighting different insider threat actors in shared homes, including the landlords as a special type threat actor outside of the home and beyond home threat actors whose involvement is specific to shared home settings. The inhabitants' level of trust fluctuates depending on the known and unknown elements of the actors involved.

\end{enumerate}

The rest of the paper is structured as follows. The next section further articulates the background and reviews related work. Section~\ref{Sec:Methodology} discusses the methodology adopted, followed by results described in Section~\ref{sec:results}. The last section discusses our study's important takeaways and limitations and gives future research directions.

\section{Related Work}
\label{Sec:RelatedWork}

There has been a robust growth of research interest in the area of multi-user smart homes focusing on different contexts, behavioural aspects, concerns, and challenges in relation to home users' S\&P perspectives. This section highlights the focal points of related research where home users' S\&P perspectives in a multi-user home were explored, primarily in traditional family home settings.

\subsection{Awareness \& Behaviours of Users}
\label{subsec:behaviour}

\subsubsection{Focus on Different Types of Users:}

Several of the studies~\cite{lin2020transferability, huang_2020_amazon, Jang2017Multi-user, lau2018alexa, zeng2017end, malkin2019privacy} focused on the influence and impact of different types of users in a multi-user home environment. Depending on their ownership, control and use of the devices, different user types were defined. \textit{Primary users} were defined as the user who owns/install/set up, and manage the computing devices at home. \textit{Secondary users} came under the categories of users who do not possess ownership/configuration control on the devices, have either very little or no control over the management, but use the devices and/or are affected by the use of such devices by other users. Researchers used various different terms to describe such users, such as \textit{bystanders}~\cite{Bernd2022MU, yao2019privacy, Ahmad_2020_TangibleMuser, Cobb_2021_MultiUser, Markey2022RolesMatter, Markey_2020_Idont_MultiUser, Markey_2020_YouJust_MultiUser, Windl2022MU}, \textit{incidental users}~\cite{moh2022characterizing, Cobb_2021_MultiUser, zeng2017end}, and \textit{passenger users}~\cite{Koshey2021MU}. Some studies~\cite{Choe2012SensorProxies, Jang2017Multi-user, lau2018alexa, ParkLim_2020_UserExpectations} examined both primary and secondary users' points of view.
 
\subsubsection{Power Dynamics between Different Types of Users:}

\citet{MackayMiller2021AbuseVictim} focused on the extreme cases of power imbalance in a multi-user home. They described how the abusers in an abusive relationship use different smart devices and try to control the home environment, i.e., temperature, hot water, and lighting, to exert their power on the other individual(s). They explained the monitoring behaviour of the abusers by using audio recording or camera surveillance and spoke about a `technocratic attitude' of the abuser, who, according to their data, is predominantly male, ignores the victim's discomfort, and follows a pattern of increasing technology use.

Unlike studies on controlling abuser behaviour and that of controlled victims~\cite{Leitao2019IPA, MackayMiller2021AbuseVictim, slupska2021threat}, most research~\cite{Koshey2021MU, Ahmad_2020_TangibleMuser, geeng2019Multi_User, yao2019privacy, malkin2019privacy} reflected on the power imbalances by exploring the positive negotiation of main users and passive acceptance of secondary users. \citet{Koshey2021MU} used a theoretical framework (domestication theory~\cite{hirsch2003information}) and stated this relationship as a spectrum where the pilot users try to accommodate the passenger users while at the same time giving priority to their own needs. \citet{Ahmad_2020_TangibleMuser} proposed a novel concept of `tangible privacy', which is built around the framework of Altman's Privacy Regulation theory~\cite{altman1977privacy} and privacy as Contextual Integrity theory\cite{nissenbaum2004privacy}. \citet{Ahmad_2020_TangibleMuser} observed that it is natural for bystanders to follow the familiar interpersonal boundary regulation mechanism to manage their privacy with IoT devices when with other people. However, in the absence of a tangible privacy mechanism from the devices, there is a mismatch between perceived privacy and actual privacy, creating uncertainty in the negotiation process. \citet{geeng2019Multi_User} highlighted the role of smart home device drivers, presented an account of the power dynamics at play in a multi-user home while focusing on key topics of types of tensions, the reason for exacerbation of the tension and ways of mitigation techniques during different phases a smart device life and thus added temporal dimension to the analysis as well.

\citet{Bernd2022MU} explained how the power dynamics between the bystanders (i.e., nannies) and owners motivate and constrain the privacy choices of the bystanders. They explained how the presence/absence of cameras at work could motivate acceptance/refusal of a job offer or sometimes constrain their choice of leaving a job, discussing/negotiating about their job. Bystanders and residents' behaviour, their different privacy perceptions, coping strategies to protect their environments, and the kind of information shared between them were all explored in a series of related papers~\cite{Markey_2020_Idont_MultiUser, Markey_2020_YouJust_MultiUser, Markey2020AllinOne_Behavior, marky2021roles}. \citet{marky2021roles} found that residents have clearer knowledge about the data collection nature of the devices and fewer misconceptions compared to the bystanders.

\subsubsection{Shared Devices \& Consequential Actions:}

We noticed S\&P discussions in a multi-user home were based primarily on two different types of smart devices, such as devices developed to be used by all members of a household, such as smart thermostats, smart cameras, smart bell or smart locks, and devices which are owned individually but can also be shared at times such as smart speakers/hubs. Although most studies focused on specific smart devices such as smart speakers~\cite{he2018rethinking, Ahmad_2020_TangibleMuser, huang_2020_amazon, Jang2017Multi-user, lau2018alexa, yao2019privacy, zeng2019HomeUser}, smart thermostats~\cite{Koshey2021MU, MackayMiller2021AbuseVictim, Cobb_2021_MultiUser, geeng2019Multi_User} and smart cameras~\cite{Bernd2022MU, moh2022characterizing} also had a fair share in the multi-user S\&P research. \citet{malkin2019privacy} mentioned how users in a multi-user home might inadvertently share search questions, listening preferences, or other personal data with others while using smart speakers. \citet{lau2018alexa} recommended smart device companies to adopt multi-user-oriented user experience design interface. \citet{moh2022characterizing} reflected on the common S\&P behaviour in a multi-user home by characterising the accidental or intentional unauthorised use of IoT devices by the residents while being shared.

Few of the past studies investigated the tensions or concerns of sharing traditional devices such as laptops and personal computers (PCs)~\cite{lin2020transferability, Windl2022MU} along with smart devices. Among other studies, \citet{lin2020transferability} explored the area of transferability of privacy behaviour between traditional computing devices and modern smart devices to find out how individuals with set privacy habits for traditional devices tend to adopt the same behaviours for smart devices as well. Several studies reflected on the nature of sharing traditional devices such as laptops and mobile phones in multi-user contexts. \citet{matthews2016she} pointed out the messiness and varied ways of sharing a laptop or mobile phone, the commonness of sharing devices such as mobile phones and computers within family home environments, and the influence of trust and convenience. In their 2022 SoK (systematization of knowledge) paper, \citet{wu2022sok} mentioned `access to shared resources' as one of the four key S\&P behaviours in the social cyber security literature and reflected on the fact that the sharing percentage is higher for desktop computers than for laptops. Studies have explored the views of romantic partners~\cite{park2018share} and cohabiting couples~\cite{jacobs2016caring} in connection with account and device sharing. Several studies~\cite{Riju2023MobilePhone, chen2022sharing, komen2016MobileSharing} investigated the intent behind device sharing to understand the underlying reason for sharing and the implications of sharing by individual users. However, these studies lack a focus on examining users in a non-family shared home setting. Exploration of sharing habits includes strangers, acquaintances, colleagues~\cite{Riju2023MobilePhone}, guest~\cite{chen2022sharing}, and friends~\cite{komen2016MobileSharing}, mainly in a family home context. There is a lack of research focusing on S\&P behaviours of shared home users concerning traditional computing devices. A specific research focus on understanding how home users behave in different non-family shared households is, therefore, an important topic.

\subsubsection{Contextual Behaviours:}

The concept of contextual behaviour has been discussed in detail by many different researchers. \citet{Bernd2022MU} as discussed above, has explained the privacy behaviour in a multi-user home from different contextual angles of the home, work, and a parental point of view and  \citet{yao2019privacy} observed that bystanders, as opposed to owners of smart devices, faced strong contextual variations as their interface with the devices varies quite prominently depending on places they visit (i.e., friends home as visitors, employees like nannies at their employer's house or even user in an Airbnb place) and the time period they spent time on such devices. \citet{Windl2022MU} expressed similar views to \citet{yao2019privacy} when they observed that bystanders privacy behaviour/concerns changed depending on the context of the device in use. \citet{Ahmad_2020_TangibleMuser} showed that contextual integrity shapes the privacy perception of the user about the devices in use and privacy violation occurs when there is a mismatch of expectation of device function/behaviour against the actual data collected by the device. \citet{he2018rethinking} presented different contextual factors such as time of the day, location of a device, and age of the person, which decide the capabilities of a device being used.

\subsection{Multi-user Home: Issues \& Concerns}

User behaviour and concerns are, to a certain extent, intertwined with each other in such a way that discussing one invariably refers to the other. While discussing the camera-related concerns for nannies about how the cameras spy on them or use illegitimately to record them, it was inevitable to discuss the S\&P behaviour of the owners as well~\cite{bernd2020bystanders,Bernd2022MU} as it was the behaviour of the owner which bought on the said concerns. \citet{Bernd2022MU} reflected on the complex contextual bases of home and workplace where the nannies have the challenges of balancing their own privacy expectations, their limited ability to make choices or express preference and that of the existing parental/employer/homeowner prerogatives. Other studies~\cite{Cobb2021MU} have found that device owners are often willing to accommodate the incident users, but incidental users have varying degrees of concern and frustration depending on the type of devices.

Smart device-related concerns were very prominent amongst the IPA victims~\cite{Leitao2019IPA}, who were concerned about the shared nature of smart devices, providing the possibilities of shared accounts, log views, and remote surveillance access by the abusers. As mentioned in the sub-section~\ref{subsec:behaviour}, \citet{Windl2022MU} highlighted the concept of a `skewed privacy concern' where the bystanders considered the laptops and personal computers significantly less privacy-concerning than the smart devices owned by the owners.

With the increasing facilities of sharing single, smart devices with multiple people, access control becomes one of the most important challenges in a multi-user home. Many of the multi-user studies~\cite{he2018rethinking, Jang2017Multi-user, zeng2019HomeUser, geeng2019Multi_User} have explored the area to discuss the level of granularity, the device-centric model, contextual dependencies, flexibility, user agency, balancing usability and complexities, etc. in relation to the topic of access control. \citet{zeng2019HomeUser} proposed different types of access control systems such as role-based, location-based, supervisory, and reactive based on the device capabilities and other factors involved, whereas \citet{he2018rethinking} added different contextual factors such as age, location of the device to the device capabilities and user relationship mix, suggesting these would better capture the user preferences and hence should guide the granular access control policy rather than the device itself. \citet{Jang2017Multi-user} used various scenarios to explain how access to information in a smart device could be categorized as high/medium or low risk depending on the device and the user and recommended the need for fine-grained access control.

\subsubsection{Insider Threats and Adversarial Settings:}

The intensity of the issues and concerns in a multi-user home is examined and reflected in a variety of spectra, starting from mild annoyance/conflict between the home users to the extremes of intimate partner violence (IPV) facilitated by smart devices and digital technologies. \citet{he2018rethinking} pointed out the scarcity of research in identifying and discussing internal threats. Research on adversarial settings at home and \textit{insider threats} has been touched upon by few studies to mainly point out the glaring absence of research in this area~\cite{slupska2019safe}. Many studies~\cite{Leitao2019IPA, MackayMiller2021AbuseVictim, slupska2021threat} have looked into how the smart devices facilitated homes, enabled the abusive behaviour inside the home, and discussed in detail the victims' concerns and the surrounding issues. On the other end of this spectrum, there lie several reasons, such as wilful disobedience~\cite{he2018rethinking} to exacerbation of existing power driver dynamics~\cite{geeng2019Multi_User} or unintentional access denial~\cite{zeng2017end}, for the inhabitants of a household which might be exhibited in mild threat-like behaviour inside a smart home.

Our study endeavours to shed light on this area, keeping in mind that inter-user relationships in a shared home are different from those in a family home, and the power dynamics in such homes are not similar, affected by many different factors and contexts.

\subsection{Multi-user Home: Coping Strategies}

Several studies discussed the mitigation techniques either by reflecting on how users manage, protect, and prevent S\&P violations in a multi-user home or by suggesting/recommending ways to protect. \citet{huang_2020_amazon} discussed how the users follow either an `Avoidance' or an `Acceptance' strategy to cope with concerns from other members of the households or visitors. Various coping strategies were mentioned by~\citet{yao2019privacy} while they explored the concept in relation to both the bystanders and the owners. Potential techniques that the bystanders wanted to use included switching off the device, deleting/jamming the data collection, adapting to what is available, or no strategy at all. Owners, on the other hand, considered the placement of the device and selective assignment of information as their coping strategies. Conversation and negotiation as a strategy was mentioned in many different studies~\cite{Ahmad_2020_TangibleMuser, geeng2019Multi_User, yao2019privacy}. \citet{Ahmad_2020_TangibleMuser} study result on coping mechanism was reflected in line with the categorisation of \citet{altman1977privacy}, i.e., `filtering', `ignoring', `blocking', `withdrawal', and `aggression'. Privacy resignation is a common theme that appeared in many of the related studies~\cite{lau2018alexa, Bernd2022MU, Markey_2020_Idont_MultiUser}.

The majority of the above discussions floated around the traditional family home settings, reflecting on the relationship between spouses, parents and children, friends and visitors to family, or employees working for the family. Some of these research data~\cite{geeng2019Multi_User, huang_2020_amazon, lau2018alexa, zeng2019HomeUser, Ahmad_2020_TangibleMuser} did include shared home residents such as the roommates in their datasets, but none of the studies specifically focused on shared home settings to understand the S\&P behaviour and attitude of the inhabitants that might be peculiar to the occupants in such environments. More often than not, the people living in shared homes are often not connected by familial bonds, and hence, the ties, trust, understanding, and obligations between the members of a shared home are different in comparison to a family home. Thus, it is important to study whether these differences are reflected in their S\&P behaviour and attitude and whether they give rise to a different set of concerns and issues. 

\subsection{Research Gaps Identified}

The above-summarised literature demonstrates a wide variety of research has been conducted on users in multi-user homes and their perception of S\&P-related issues and concerns. However, the context of research, as seen, has primarily been on users of family homes. Several past studies~\cite{zeng2019HomeUser, huang_2020_amazon, Jang2017Multi-user, Markey_2020_Idont_MultiUser} did include users of non-family homes in their analysis \textcolor{black}{to evidence the varied multi-user home settings. Focusing on access control issues in a multi-user home, \citet{zeng2019HomeUser} commented on roommates' respect for each other's space. \citet{huang_2020_amazon}, as part of their discussions on users' concerns and coping strategies on shared smart speakers in multi-user homes, included some data from non-family households reflecting on trusting issues and how roommates in a multi-user scenario use shared smart speakers as coping strategies. While discussing the challenges of sharing smart devices in a multi-user household, \citet{Jang2017Multi-user} highlighted a typical non-family household scenario and the possibilities of having alternate primary users. \citet{geeng2019Multi_User}, who focused on understanding the tensions and cooperation among users of a multi-user home during different phases of smart device use, expressed their difficulties in being able to recruit a large number of non-family households.}

\textcolor{black}{Although some past studies considered non-family household scenarios, they were always researched and presented as `\textit{part of}' a bigger picture, i.e., a multi-user household with family and non-family rather than studied `\textit{holistically}' to understand the S\&P nuances of the users living in these settings as a separate group. The specific S\&P impact and implications that a non-family home context as a whole might exhibit are certainly worth more investigation.} As mentioned in Section~\ref{Sec:Intro}, this type of shared home would possibly always have multiple primary users for shared devices, several passive users of devices which might be bought by other users in the house, constant possibility of exposure of their devices to outsiders, i.e., friends of friends with whom they might not be familiar with and/or probable past/ex-users who might be privy to the detail S\&P settings if appropriate care has not been taken. The current research aims to fill such research gaps.

\section{Methodology}
\label{Sec:Methodology}

As reflected from the research questions described in the Introduction section, our study aims to understand the S\&P perspectives, including awareness, behaviours, and concerns of co-habitants in shared home environments. We decided to collect our data from two different sources: data collected via an online survey and user-generated content (UGC) from an online platform. There were several reasons behind the decision to use the two different sources of data. First, usability security and privacy (USP) studies generally use traditional empirical methodologies, i.e., surveys and interviews. Since our study falls under such a category, we chose online surveys as our first choice of data collection. Second, we know that survey responses can be biased by the questions of the researchers and also by the survey respondents' desire to submit an answer that fits the norm. We, therefore, decided to collect additional data from a real-world online platform, which was collected in a passive manner so that there was no influence of the researchers~\cite{spiti2022social, rocha2023passive}. We wanted to find out whether there are any S\&P-related online discussions with regard to the shared home environment and, if there are some, to investigate the nature of such discussions. Throughout this paper, whenever we use the term `dataset', it will include both the survey data and the data collected from the online platform.

\subsection{Collecting Online Data}

Among multiple online platforms, we decided to choose Reddit. There were a few reasons behind this choice. 1) Reddit is a highly popular online forum with 52 million daily active users, active on over 3.5 million different communities called subreddits~\cite{Bleu2023RedditStatistics}, each devoted to discussing a different subject. So, the probability of getting relevant data on our chosen topic was high. 2) Reddit is one of a few online platforms that have been actively studied~\cite{proferes2021SLR-Reddit} by many researchers due to the richness of UGC~\cite{PARK2018RedditUse, Wang2023RedditUse}. 3) Past research~\cite{jamnik2019use} has shown that Reddit is a valuable platform for getting high-quality data from a diverse population with good measurement reliability inexpensively.

To collect online posts relevant to our research questions, we first needed to select one or more relevant subreddits related to S\&P aspects of home users and then relevant Reddit posts on selected subreddits. To this end, we first searched on Reddit using the keyword ``Home Security'' to find relevant subreddits. Out of the 56 returned subreddits, we selected seven that meet the following criteria: 1) the description is pertinent to computing devices and/or networks used in a home setting, 2) there were at least more than 1,000 members, and 3) the topics covered are not very technical in nature, e.g., `r/cybersecurity' focusing on technical security discussions was excluded. In addition, we decided to add three popular subreddits related to smart speakers (Google Home, Smart Things, and Amazon Echo), drawing on the evidence from past studies where we noticed many past studies~\cite{Ahmad_2020_TangibleMuser, geeng2019Multi_User, he2018rethinking, huang_2020_amazon, Jang2017Multi-user, lau2018alexa, yao2019privacy, zeng2019HomeUser} that used one or more of the three types of smart speakers to demonstrate S\&P perspectives of users in multi-user homes. The ten selected subreddits include `r/technology', `r/privacy', `r/smarthome', `r/homeassistant', `r/homenetworking', `r/homeautomation', `r/homesecurity', `r/googlehome', `r/Smartthings', and `r/amazonecho'.

In order to collect relevant Reddit posts, we designed a search query with a selection of keywords. Our keyword selection was guided by the research questions and the context of our study. The research questions focus on two main areas, i.e., S\&P perspectives of home users and the shared home context. As mentioned above, the selection of subreddits focused on home security-related topics to satisfy the criteria of S\&P perspectives of home users. Next, we needed to find relevant posts that reflect the shared home context. We, therefore, chose the following list of keywords that are synonymous with the meaning of shared home and might be used by Reddit users in such a context. These keywords were chosen after a detailed discussion between the authors of this paper. In the following search query, the character `|' represents the Boolean operator OR:
\begin{center}
\textit{("shared apartment" | housemate | "shared accommodation" | "shared room" | "shared flat" | "Paying guest" | "Live-in landlord" | "Hospital accommodation" | "shared house" | communal | hostel | hospice | "communal home" | "student accommodation" | "staff accommodation" | flatmate | roommate)}.
\end{center}

We used the Pushshift API (\url{https://github.com/pushshift/api}) to search for and extract the relevant data dated between 1 May 2021 and 1 May 2023 from the selected subreddits, leading to 411 posts related to shared home environments. Pushshift API ingests Reddit data through its official API as per Reddit's data collection and maintenance terms of service~\cite{Reddit2023Compliance} and makes it available for public use. We manually analysed each post to filter posts that are S\&P related (58) and discarded posts that seemed, in our expert opinion, to be posts submitted by an expert user. We finally got 46 posts for further analysis.

\subsection{The Online Survey}

The survey included in the Appendix was used to understand the behaviours, attitudes and concerns of co-habitants in shared home environments. Some of the questions in the survey followed the topics we discovered earlier in Section~\ref{Sec:RelatedWork} and were included to comprehend any pronounced similarities or differences in the S\&P perspectives of the people in shared homes, as opposed to the traditional family homes. Although questions in the survey cover previously studied topics in past research, we constructed all our survey questions independently for our work.

The questions in the survey formed three different sections: collecting demographic data in a multi-user home, S\&P behavioural data related to network and router use, and finally, the S\&P behaviour and concerns with regard to devices and other users in the shared home environments. Two researchers rigorously examined and validated the survey questions in several sessions. A pilot survey was conducted with participants recruited using Prolific (\url{https://www.prolific.co/}), an online participant recruitment system used widely by many researchers~\cite{Turner2022ProlificStudy, Windl2022MU, marky2022PrivacyTea, Cobb2021MU}. The survey questions were edited after reviewing the pilot survey data results. The final survey was also conducted using Prolific. 174 people participated in the final survey, each taking an average of 8 minutes to complete the survey. Participants were rewarded for their time at a payment rate of £11.10 per hour. The authors' institution's research ethics committee gave a favourable opinion of the study. After carefully examining the data, we rejected 25 participants because of the incompleteness of data, leading to 149 valid participants.

Our survey questions were designed to understand several different concepts, including identifying peoples' device use, access and management of shared home networks, and residents' trust level towards their cohabitants' visitors and endeavoured to discern various threat actors specific to shared home users. We calculated the descriptive statistics of the survey data and conducted some inferential statistics to explore the associations between participants' perceived S\&P risks in shared home environments and the factor that might be responsible for the perceived risks, such as potential malicious behaviours from other co-habitants and their visitors.

\subsection{Analysis of data}

We used both descriptive and inferential statistics to analyse the survey and online data quantitatively. All analyses were conducted using IBM SPSS Statistics (V.29.0). The descriptive statistics were primarily conducted to understand the frequencies of survey question responses. A series of chi-square tests were performed on the online survey data as part of the inferential statistics to understand whether responses to certain questions may be associated with other responses. Specifically, we wanted to find out whether the participants' S\&P concerns are related to how their perception of S\&P-related risks in the shared home environment and their understanding of S\&P risks from sharing computing devices amongst the members of the shared home.

The Reddit data was also qualitatively analysed. We open-coded each of the Reddit posts to understand the different messages within them. Following the interpretative phenomenological analysis (IPA) approach~\cite{alase2017interpretative}, we decided on sub-themes for each post and then compared the individual sub-themes to decide on the \textcolor{black}{high-level} theme of all the Reddit posts. The first author performed open coding to develop a codebook capturing the main themes, while the second author of the study used that same codebook to code the Reddit posts independently, adding/modifying the codes as they felt fit. The second meeting decided on the axial codes, that is, the second level of codes, which identifies the emerging themes~\cite{williams2019art} and the possible final themes. \textcolor{black}{Our main aim was not to see whether the same open codes were chosen for each post but whether both coders linked the same posts to the same axial codes.} Subsequent meetings discussed any modifications and/or additions to the codebook and the major themes. The posts were coded with the new coding scheme. We aimed to achieve a richer interpretation rather than a consensus of meaning, as pointed out by~\citet{braun2019thematic}. We used a research software system called MAXQDA (\url{http://www.MAXQDA.com/}) to help our qualitative analysis. To maintain the anonymity of Reddit users, we have paraphrased all the quotes from the users throughout this article.

\section{Results}
\label{sec:results}

This section begins by describing some descriptive statistics of our survey data. It then shows results from both the online survey and analyses of the Reddit data, organised around the two research questions.

\subsection{Survey Participants}

Table~\ref{tab:survey_demographics} shows the demographic details of our survey participants. The average size of the household is 4.27, with 25.3\% living in a 4-person household, 24.7\% in a 3-person household, and 11\% in a 5-person household. For eight households, the number of co-habitants reaches up to 8, and for two households, the number of co-habitants is 20 and 30, respectively. Nearly half (50.3\%) of participants belong to the age group of 26--35, and the second largest age group is 18--25 (23.5\%).

\begin{table}[!htb]
\centering
\caption{Participant demographics ($n=149$)}
\label{tab:survey_demographics}
\begin{tabularx}{\linewidth}{rX}
\toprule
\textbf{Demographic} & \textbf{Statistics}\\
\midrule
\textbf{Gender} & Male (54.4\%); Female (45.6\%); Others (0.0\%)\\
\textbf{Age range} & 18--25 (23.5\%); 26--35 (50.3\%); 36--45 (16.8\%); 46--60 (9.4\%); $>60$ (0.0\%)\\
\textbf{Type of shared accommodation} & Privately rented (88.0\%); Student accommodation (8.7\%); Staff accommodation (1.3\%); Hostel (1.3\%)\\
\textbf{Student status} & Yes (29.3\%); No (70.7\%)\\
\textbf{Employement status} & Full time (70.0\%); Part time (20.0\%); Unemployed (10.0\%)\\
\textbf{First degree or above (ICT-related)} & Yes (26.1\%); No (including education below the first degree) (73.9\%)\\
\textbf{ICT-related job} & Yes (24.2\%); No (including unemployed) (75.8\%)\\
\bottomrule
\end{tabularx}
\end{table}

The following chart shows the number of people living at each participant's home. We assume people staying with more than 20 or more participants to be living in student/staff/hostel accommodation. We can deduce from this figure that more than 50\% of households live with $\geq 4$ inhabitants.

\begin{figure}[!htb]
\centering
\includegraphics[width=0.8\linewidth]{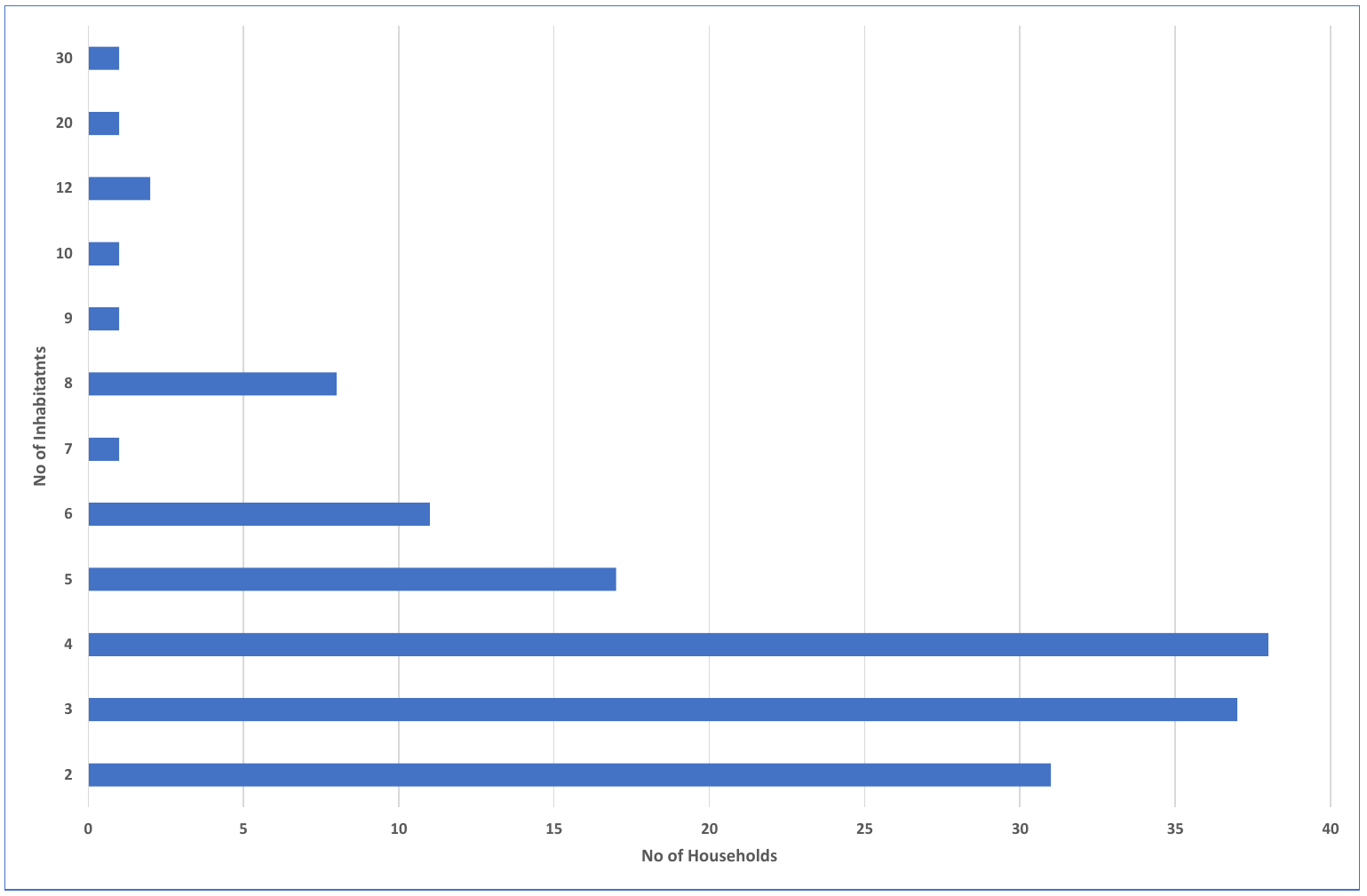}
\caption{The distribution of the number of inhabitants per household}
\label{fig:NoOfUsers}
\end{figure}

\subsection{Device Ownership \& Sharing}

In shared home environments, who own(s) and manage(s) what devices can be complicated due to the existence of multiple co-habitants without a familial bond. We observed some interesting points with regard to the spread of devices in our dataset. As highlighted in Table~\ref{tab:DeviceSpread}, over half of most types of reported devices were owned/managed by the landlords. One possible explanation for such a phenomenon is that most shared homes are often fully or partially furnished; therefore, the landlord is the natural party who is expected to provide basic home appliances and computing devices. This is reflected in another observation that 89\% of our survey participants lived in privately rented households. Note that the landlord-tenant relationships in terms of device ownership and management can also exist in more traditional family homes as long as the home is privately rented. However, as far as we know, this particular aspect in relation to multi-user S\&P challenges has not been researched in past studies for any home environments, so our work is the first one exploring it. Considering the nature of most home devices, we can assume the actual residents are the frequent users of such landlord-provided devices; however, our survey data show that our participants were not provided much information on the extent of data collection/storage of these devices from their landlord: 52.3\% of them did not receive any details regarding the devices, and only 22.1\% received some functional and/or operational details such as WiFi password and how to use the smart thermostat. A survey question about the sharing habit of devices revealed that 27.5\% of the residents shared their smart speakers. When asked about their behaviours on sharing traditional computing devices, a surprising 24.7\% of the survey participants mentioned that they regularly shared their laptops/PCs/tablets with other co-habitants and 5.3\% of the participants even shared their smartphones.

Most device-related posts on Reddit are about investigating secure ways of owning and/or sharing devices within a shared home. Sometimes, the poster wanted to have control of a device so that they could control the activities of the other co-habitants, and sometimes, the poster wanted to understand whether their co-habitants' devices were spying on them. There are five posts where the landlord-resident relationship was mentioned. In all of these cases, the router was owned by the landlord. Although the landlord was not involved in any of the S\&P concerns mentioned in these posts, they had the controlling say in how these issues were managed (i.e., in four out of those five cases, the posters referred to the landlord to solve their problem).

\begin{table}[!htb]
\centering
\caption{Statistics of the device ownership and control}
\label{tab:DeviceSpread}
\begin{tabularx}{\linewidth}{Xcccccc}
\toprule
\textbf{Type of Devices (owned/managed by)} & \textbf{Landlord} & \textbf{Participant} & \makecell{\textbf{Single}\\\textbf{Co-habitant}} & \makecell{\textbf{Multiple}\\\textbf{Co-habitants}} & \textbf{\#(Homes)}\\
\midrule
\textbf{Router} & \textbf{51.0\%} & 27.9\% & 9.5\% & 11.6\% & 148\\
\textbf{Smart Thermostat} & \textbf{80.3\%} & 8.5\% & 7.0\% & 4.2\% & 76\\
\textbf{Smart Lock} & \textbf{71.0\%} & 12.9\% & 12.9\%& 3.2\% & 36\\
\textbf{Smart Speaker} & 12.2\% & 43.3\% & 23.3\% & 21.1\% & 90\\
\textbf{Security Camera (external)} & \textbf{76.6\%} & 12.8\% & 4.3\% & 6.4\% & 47\\
\textbf{Security Camera (internal)} & \textbf{50.0\%} & 26.7\% & 16.7\% & 6.7\% & 30\\
\textbf{Smart Bell} & \textbf{56.8\%} & 16.2\% & 16.2\% & 10.8\% & 37\\
\textbf{Smart Fridge/Freezer} & \textbf{50.0\%} & 16.7\% & 25.0\% & 8.3\% & 12\\
\textbf{Smart TV} & 16.0\% & 49.6\% & 20.8\% & 13.6\% & 125\\
\textbf{Smart Meter} & \textbf{69.4\%} & 12.5\% & 6.9\% & 11.1\% & 72\\
\bottomrule
\end{tabularx}
\end{table}

\subsection{RQ1: What are the S\&P awareness and behaviours of the people living in a shared home setting?}
\label{subsec:RQ1}

\subsubsection{\textbf{Types of Users -- Contextual Roles}:}
\label{subsec:UserType}

As shown in the related work, there are different types of user relationships at play in a multi-user home environment, including primary and secondary users or pilot and passenger users~\cite{lau2018alexa, malkin2019privacy, zeng2017end, lin2020transferability, huang_2020_amazon, ParkLim_2020_UserExpectations, Koshey2021MU}, bystanders~\cite{Bernd2022MU, yao2019privacy, Ahmad_2020_TangibleMuser, Markey2022RolesMatter, Windl2022MU} and incidental users~\cite{moh2022characterizing, Cobb2021MU}. We examined different scenarios in our Reddit data and the survey data to find out what user types existed in shared home settings. We observed that shared home users exhibit roles similar to those in traditional family homes. However, the nature of such user roles often differs, reflecting different associations amongst co-habitants, which are mostly horizontal in nature due to the lack of a hierarchical familial structure as in traditional family homes. For example, co-habitants in a shared home normally have an equal right to use landlord-provided devices, which results in a \textit{primary-primary} user relationship rather than the more typical \textit{primary-secondary} user relationship seen in a family home. A bystander in a family home is always portrayed to represent a temporary visitor, i.e., visitors coming over and domestic workers such as nannies working at home, as opposed to in a shared home where other co-habitants can become bystanders and they are not usually temporary visitors to the home. Both our survey data and the Reddit posts evidenced the existence of these above relationships. A detailed examination of our findings revealed four different types of users in addition to other types already described in past studies. The description below endeavours to define such users.

\begin{enumerate}
\item \textbf{ExternalPrimary}: These users have primary ownership and control of the devices being used but do not reside inside the shared home. The main example of this kind of user is the landlord of a shared home who does not live in the home but controls one or more devices inside the home. For example, the landlord commonly owns and controls the router in a rented home, so they are the primary users, but in most cases, they do not live inside the shared home. Hence, we labelled them as ExternalPrimary.

\item \textbf{InternalPrimary}: These are users who have their own devices and control them as administrators. For example, many shared home users would have their own smart speaker or security camera, which they own and control themselves. They are the primary users of these devices, and they live inside the shared home. These users are, therefore, termed as InternalPrimary users. 

\item \textbf{ActiveSecondary}: These are users who do not have primary control over one or more devices in a shared home but actively use them. For example, when some of the shared home users do actively use a smart speaker owned by another co-habitant, they are called secondary users as they do not have the primary ownership or admin rights over the device, but at the same time, they use the devices actively so are labelled as ActiveSecondary.

\item \textbf{PassiveSecondary}: These users do not have primary ownership or admin control over one or more devices. Additionally, they do not actively use such devices, either, however, they can be affected by other co-habitant(s)'s use of such devices. For example, when a user of a shared home installs a security camera in front of their room in a common corridor, the movements of other co-habitants get recorded whenever they pass by the corridor or do any activity there. The latter users are not actively using the camera, but their privacy could be affected by the existence of such a camera. Such users are named in our study as `PassiveSecondary' users. The nature of these users simulates the same features as a \textit{bystander} as mentioned in past studies~\cite{bernd2020bystanders}, with the only difference being the following: \citet{bernd2020bystanders}'s \textit{bystanders} are users who temporarily visit the device owner(s)' home whereas our \textit{PassiveSecondary} users live in the same home.
\end{enumerate}

\begin{figure}[!htb]
\centering
\includegraphics[width=\linewidth]{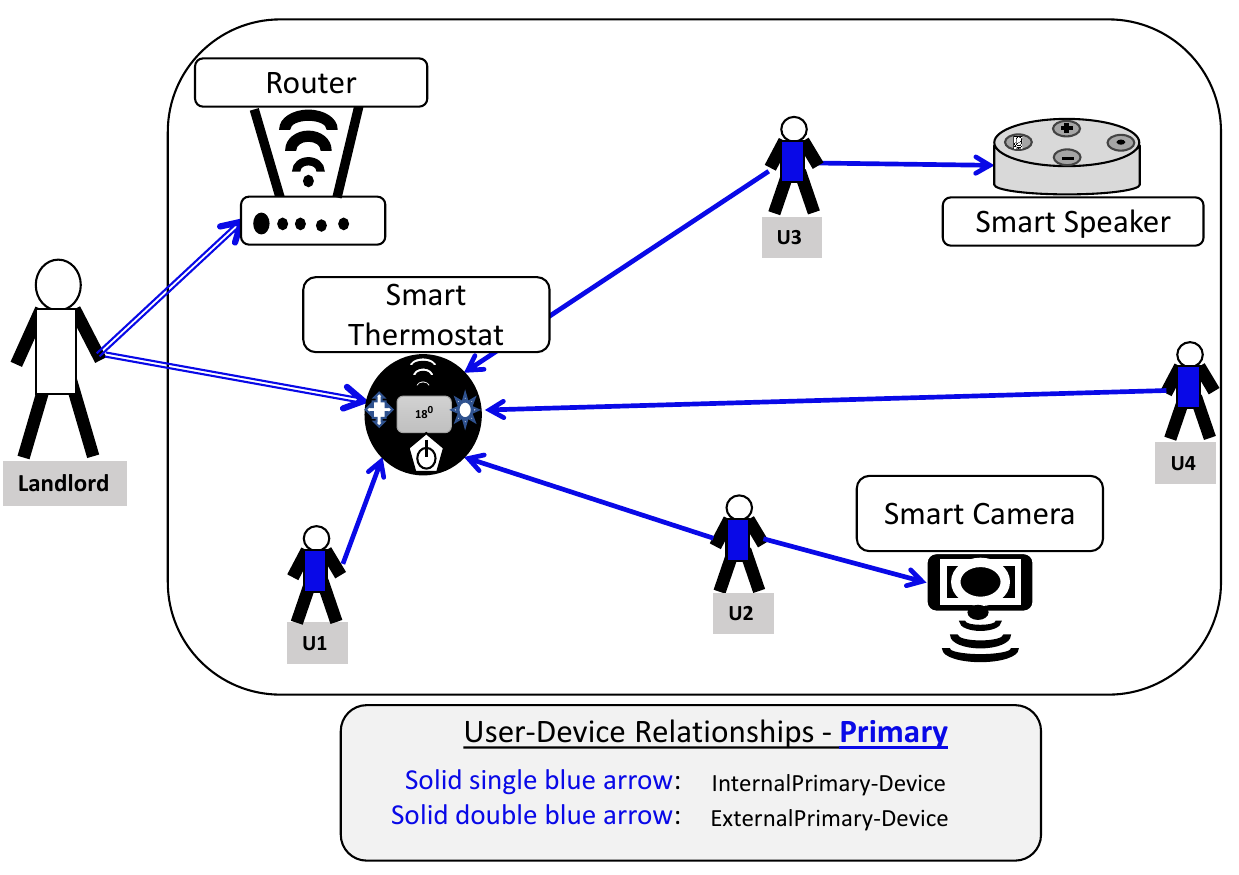}
\caption{A visual illustration of user roles within a shared home in terms of owning, managing, and using computing devices -- Primary users}
\label{fig:Primary}
\end{figure}

\begin{figure}[!htb]
\centering
\includegraphics[width=\linewidth]{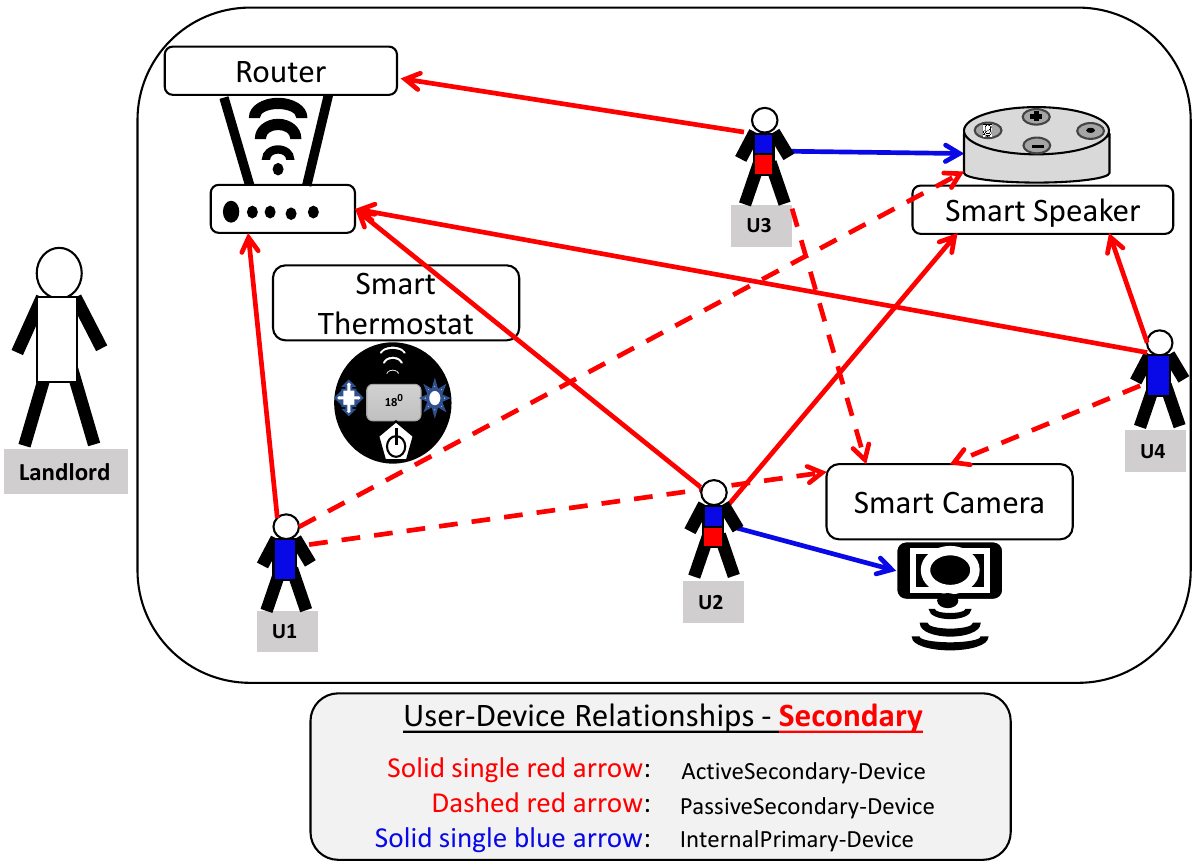}
\caption{A visual illustration of user roles within a shared home in terms of managing and using computing devices -- Secondary users.}
\label{fig:Secondary}
\end{figure}

Figures~\ref{fig:Primary} and \ref{fig:Secondary} represent an imaginary example scenario of a shared home with four co-habitants (U1 -- U4) and an externally living landlord (reflecting an average case according to our survey data) and the possible connections between these users and four home devices: 1) a router, 2) a smart thermostat, 3) a smart camera, and 4) a smart speaker. The example scenario involves four different types of users, which we describe below. To represent the different user-device relationships, we use different types of line styles and line colours in these figures, as explained below.
\begin{itemize}
\item A \textbf{solid single blue arrow} represents a relationship between an InternalPrimary user and a device. 

\item A \textbf{solid double blue arrow} represents a relationship between an ExternalPrimary user and a device.

\item A \textbf{solid single red arrow} represents a relationship between a SecondaryActive user to a device.

\item A \textbf{dashed red arrow} represents a relationship between a SecondaryPassive user and a device.
\end{itemize}

The \textbf{router} is owned and managed by the landlord, so they are the designated primary user of the device. Because, in this case, they live outside of the shared home, they are labelled as an \textbf{ExternalPrimary} user, and their relationship with the router is displayed with a solid double blue arrow in Figure~\ref{fig:Primary}. All other users, U1 -- U4, use the router actively to connect to the internet. They, therefore, are labelled as \textbf{ActiveSecondary} users, and their relationship with the router is depicted in Figure~\ref{fig:Secondary} with solid red arrows.

The landlord owns the \textbf{smart thermostat}, and they also have the administrative rights to control the device if they want to do so. They, therefore, are a primary user of the device. On account of his living outside of the house, they are labelled as an \textbf{ExternalPrimary}. His relationship to the thermostat is displayed as the solid double blue arrow in Figure~\ref{fig:Primary}. The administrative rights to control and manage the thermostat are also given to all the co-habitants U1 -- U4. Hence, all of them are considered primary users who are internal to the home and labelled as \textbf{InternalPrimary}. Their relationship with the thermostat is displayed with a solid single blue arrow in Figure~\ref{fig:Secondary}. This scenario, therefore, creates multiple primary users in the shared home context.

U3 owns and controls a \textbf{smart speaker} and, therefore, is the primary user of that speaker. This relationship is displayed in Figures~\ref{fig:Primary} and \ref{fig:Secondary} by the solid single blue arrow between U3 and the smart speaker. U3 allows U2 and U4 to use the smart speaker whenever they want. U2 and U4 are, therefore, designated as secondary users of the device who actively use the smart speaker and labelled as \textbf{ActiveSecondary}. Their relationship to the smart speaker is represented with a solid single red arrow. U1, on the other hand, could be playing the role of \textbf{PassiveSecondary} as the smart speaker might accidentally record U1's conversation as they participate in any common discussion.

U2 owns and manages an internal \textbf{smart camera} for their own use. They, therefore, are considered the primary users of that camera and designated as \textbf{InternalPrimary}. Their relationship with the camera is denoted by a solid single blue arrow. Other co-habitants are not given permission to use the camera, but the camera actively records all movements of anyone passing in front of it and is placed near the front of U2's room door, which opens to the communal corridor of the shared home. The camera, therefore, quite possibly records the movements of the other co-habitants U1, U3 and U4 whenever they are in the corridor. These other users, therefore, are passively exposed to the camera all the time and labelled as \textbf{PassiveSecondary}. In Figure~\ref{fig:Secondary}, this relationship is exhibited with a dashed red arrow.

The implications of these different relationships in a shared home and how they relate to a traditional family home are discussed further in Section~\ref{Sec:Discussion}.

\subsubsection{\textbf{S\&P Awareness and Behaviours -- Home Network}:}

38.3\% (57/149) of survey participants thought the computing network in a shared home is less secure than that in a traditional family home. Two main reasons given for this perceived lower security include reduced trust in other co-habitants and their visitors and the constant exposure of the shared home to many strangers. A similar proportion of participants (60/149, 40.27\%) thought that shared homes have a similar level of security to traditional family homes. 14.1\% of the participants were not very sure about this, whereas 7.4\% believed that shared homes are less secure when they moved in but felt secure after settling down. We observed that the landlord provided 51.0\% of the households' routers, but 91.3\% of the participants shared a WiFi password, using a router that was bought by themselves (30.9\%) or another co-habitant (10.1\%). 36.9\% of the survey participants (55/149) had administrative rights on the router to change administrative settings; however, not everyone was engaged in administrative activities. From those participants who had administrative rights to make changes in the WiFi, we noticed that some co-habitants were familiar with changing the WiFi password (30.9, 17/55) or checking the status of devices connected to the WiFi network (40.0\%, 22/55) and a few also ventured into setting up a guest network (9.0\%, 5/55). Interestingly, 14.0\% (21/149) of the survey participants did not know or care about the administrative rights of the routers they used. Posts in our Reddit data also demonstrated shared home users' understanding of the requirements of a secured home and their inclinations towards building more secure home networks. For example, one of the posts queried about the creation of a guest network purely to create a safe boundary between different co-habitants of a shared home so that their devices do not get infected in case of other co-habitants use a compromised device in the home.

Exploring some of the other S\&P behaviours of shared home users, our survey discovered that 79.2\% (118/149) of the participants shared their WiFi password with all visitors to their homes, and the shared password had never changed afterwards. 60.4\% (90/149) of the participants reported that their router had been placed in a common area accessible to everyone who would visit the house. Although 53.6\% (80/149) of the participants were aware of the fact that WiFi can be spied on, 28.2\% (42/149) did not think about it, and 6.0\% (9/149) did not care about it. Additionally, 12.1\% (18/149) of the participants were unaware of it. In order to find out whether these behaviours have any statistically significant association with our survey participants' opinions about the lack of security in comparison to a traditional family home, we decided to perform a number of $\chi^2$ tests. With $\alpha = .05$, our results indicated a statistically significant association ($\chi^2 (1, n=149) = 39.916, p = .002$) between our participants' perception of security in a shared home network and their awareness of being spied on in the network. We also found a strong association ($\chi^2(1, n=149) = 50.634, p = .001$) between our participants' reported behaviours of sharing their WiFi password with visitors of other co-inhabitants and their perception of shared home security. Table~\ref{tab:NetworkConcern} presented later in Section~\ref{subsec:RQ2} reveals more associations on this above assertion.

We observed that some Reddit posters used terminologies such as \textit{`extremely paranoid and worried to share the WiFi connection'} with their roommates or expressed the desire to learn to separate their network from their co-habitants so that they could protect their security and privacy better. There were 32.6\% (15/46) network security-related queries in our Reddit data, and almost all of them were posted regarding secure configurations and settings of the network.

\subsubsection{\textbf{S\&P Awareness and Behaviours -- Around Devices Users}:}

As evident from the survey data, 75.8\% (113/149) of our participants used and shared devices that were provided by the landlord or the property owner/manager. Apart from the landlord-provided devices, some participants also shared personal devices with other co-habitants in the shared home. 24.8\% (37/149) shared their laptop/PC/tablet with other co-habitants. Only 12.8\% (19/149) were concerned that co-habitants might spy on them using the shared computing devices, and only a small percentage of 4.7\% (7/149) had some problems such as malware infection or illegal downloading after they shared their devices with their co-habitants. 38.23\% (57/149) participants of our survey shared their smart speakers with other co-habitants, and only 5.4\% (8/149) were concerned about any privacy violation from their co-habitants. 28.8\% (43/149) mentioned that the smart speaker was kept in a publicly accessible place for anyone to access it, and only 8.7\% (13/149) of the participants had set individual voice profiles on the hub so that sensitive data could not be easily accessible by untrusted persons. The above data give a slight indication that shared home users might be more concerned about sharing their traditional devices than their smart devices. We observed that more than half of the participants (56.9\%, 84/149) actively tried to move their devices, such as smart speakers, laptops, PCs, and tablets, to a secure place before any parties and other get-together events happened in their shared homes. For 44.9\% (67/149) of the participants, there was a smart camera or a smart doorbell in the shared home; however, 20.8\% (14/67) of them were unaware of how and where the recordings from such a device are kept, and for 25.4\% (17/67) the recordings were stored under the control of the landlord. Most participants exhibited good S\&P behaviours around portable computing devices such as laptops by switching them off and keeping them in secure places when not used (57.7\%, 86/149) and locking these devices when they leave the device temporarily (59.7\%, 89/149). These statistics seemed to indicate that residents in shared homes are likely more aware of S\&P issues with regard to traditional computing devices such as laptops.

Analysis of Reddit data exhibited different S\&P characteristics of shared home users. At times, the poster wanted to learn whether sharing a particular smart device is secure, and at other times, the poster sought advice or information on specific security settings on devices or the purchase of specific smart devices to prevent privacy violations. For example, one of the Reddit posters wanted to connect a smart bulb with a motion detector so as to deter their roommates from accessing personal belongings and information, while another poster wanted to know the ways of restricting Amazon Alexa devices from sharing selected information. Out of the 46 Reddit S\&P-related posts that we analysed in detail, 67.4\% (31/46) are on smart devices. We noticed only two occurrences of posts with regard to traditional devices (PCs and laptops).

\subsubsection{\textbf{S\&P Behaviours -- Around Other Users}:}

Researchers have observed that a social context should always be in the background when we analyze any S\&P behaviours (similar to any other type of human behaviours)~\cite{Das2014SocialInfluence}. Past studies have looked into the social influence on S\&P behaviours, attitudes, and practices~\cite{Das2019SocialInfluence, Watson2020MU} to understand how specific S\&P behaviours are triggered and the impacts thereof. While endeavouring to understand the nature of S\&P social interactions between co-habitants, we found our survey participants had engaged in various S\&P discussions such as placement of shared devices, sharing of passwords, creating separate guest networks with the most common form of discussion around sharing (47/149, 31.5\%) of different devices such as routers, smart locks, and security cameras. Discussions also touched upon areas such as the placement of smart routers and smart speakers (25/149 16.7\%), installing anti-virus/malware protection on personal devices to keep the network secure (25/149 16.7\%), how to maintain security and privacy of devices (23/149 15.4\%), and creating guest networks (19/149, 12.7\%). Upon questioning whether the co-habitants sought help from each other, 32.8\% (49/149) mentioned affirmatively regarding S\&P settings of various devices. There are also mentions of help in relation to S\&P problems such as cleaning infected machines, phishing( 38/149 25.5\%) and other related configurations such as setting up VPNs, creating cloud-based backups, setting up MFA, and purchasing security software (34/149 22.8\%).

We observed that Reddit posts are more in relation to seeking S\&P-related help and advice. Some of the frequent help-seeking posts are in relation to the use and purchasing of security cameras (16/46, 34.8\%), setting up secure networks (11/46, 23.9\%), and privacy in communal areas (14/46, 30.4\%). Note that there can be overlaps between the different topics, i.e., some of the security network-related posts contain messages about security cameras as well, and some posts about security cameras include privacy-related queries. In several of the posts (34/46, 73.9\%), the poster was desperate to know ways to prevent privacy risks using smart devices, which arose from conflicts with roommates. Sometimes, the behaviour displayed is conciliatory in adapting to S\&P actions of other co-habitants without any questions, and at times, the poster expressed frustration and sought ways to modify the situation. In one of the cases, the poster wanted to get advice on persuading their roommate not to buy a specific brand of vacuum cleaner that they believed was privacy-invasive. At times, tech-savvy residents undertook the responsibility of technical and S\&P work inside the shared home but did not possess the patience to explain to other co-habitants. We noticed this happening in one of the messages where the user felt frustrated because their co-habitant disabled and modified advanced settings and did not explain the reasoning behind these actions. This left the poster annoyed and being less aware of the S\&P situation of the shared home.

To summarise, we observed that the majority of shared home co-habitants consider a shared environment as less secure than a traditional family home, which could be because they felt they had to share their WiFi password with less trusted people and their fear of being spied upon by such people. We found that existing S\&P-related relationships between co-habitants in a shared home could be very complicated depending on the devices used. This sometimes led to situations with multiple/all device users having primary access to the device, a situation that is not very common in traditional family homes. Our findings also revealed the unique S\&P-related role the landlord plays in a shared home environment, where they can be considered outside ``insiders''. Note that such a role can also exist in traditional family homes when a whole family lives in a rented home from a landlord. This observation raises the point that any future research and discussions regarding S\&P of the shared and traditional family home environments should seriously include landlords in its scope. We noticed that many participants endeavoured to behave securely by having various S\&P discussions with other co-habitants within their shared homes and sought S\&P-related advice when needed. However, their efforts to be secure in a multi-user shared home environment were limited by a lack of knowledge and skills and also by how cordially the other co-habitants behaved.

\subsection{RQ2: What S\&P concerns are experienced by shared home co-habitants, and which threat actors are the likely sources of these concerns?}
\label{subsec:RQ2}

This section analyses the S\&P concerns that were expressed in the Reddit posts we collected and by the participants of our survey. We investigated the origin of the threats in our dataset and drew possible threat actors who might be responsible for the concerns of the co-habitants.

\subsubsection{\textbf{S\&P Concerns: Network, Device, Co-habitants and landlords}:}

As mentioned in Section~\ref{subsec:RQ1}, according to our survey data, 38.9\% (58/149) of the participants believed the shared home network is less secure than a traditional family home, and 50.3\% (75/149) thought that devices in a shared home are less secure than those in a family home. 88.6\% (132/149) of our survey respondents were from privately rented flats with multiple tenants, which, as shown by our further examination, is one of the reasons why users of these types of houses perceived a higher level of S\&P risks than in a family home environment. We found that some participants were sometimes concerned about threats posed by their co-habitants (14.0\%, 21/149) and more about visitors of their co-habitants (23.4\%, 35/149).

\textcolor{black}{There are two main questions in our survey, B7 and C11 (see the survey questionnaire in the Appendix), which asked the participants about their S\&P concerns in living in a shared home network with shared devices compared to a traditional family home. Intending to find any possible associations between the S\&P concerns of the participants with various factors such as the nature of the shared home, device ownership, S\&P behaviours and perceived risks of co-habitants and the landlord, we performed a series of Pearson $\chi^2$ tests ($n=149, \alpha = .05$). Tables~\ref{tab:NetworkConcern} (tests pertaining to co-habitants' concerns in a shared home network) and \ref{tab:DeviceConcern} (tests pertaining to devices used in a shared home) show the results of those tests.} These results include only the reasons with a significant effect. As can be seen, users probably formed their perception of security in a shared home for many different reasons, ranging from threats posed by other co-habitants or visitors of other co-habitants to their fear of devices being tampered with when unattended. \textcolor{black}{One important finding is about our survey participants' S\&P concerns due to the fact that some devices used are owned and managed by the landlord and that the tenants did not receive any relevant documentary information about such devices. We have highlighted these associations in both tables with an asterisk at the beginning of relevant entries.}

\begin{table}[!htb]
\centering
\caption{Statistical results of Pearson $\chi^2$ tests on probable reasons of shared home co-habitants' perceived higher S\&P risks in a \textcolor{black}{\textbf{shared home network/WiFi}}}
\label{tab:NetworkConcern}
\begin{tabularx}{\linewidth}{Xccc}
\toprule
\textbf{S\&P Concern} & \textbf{$\chi^2$ Statistic} & \textbf{df} & \textbf{$p$-Value (2-sided)}\\
\midrule
Fear that WiFi can be spied upon & 40.743 & 21 & .006\\
(*) Living in a shared student accommodation & 19.115 & 7 & .008\\
(*) Ownership of the router & 68.705 & 35 & $<.001$\\
Worry that the co-habitants or their visitor might spy using shared devices & 22.039 & 7 & .003\\
Probable S\&P risks from the co-occupants & 23.299 & 7 & .002\\
Probable S\&P risks from the co-occupant's visitors & 14.629 & 7 & .041\\
Probable risks of tampering with computing devices from the co-occupants when not at home & 14.187 & 7 & .048\\
\bottomrule
\end{tabularx}
\end{table}

\begin{table}[!htb]
\centering
\caption{Statistical results of Pearson $\chi^2$ tests on probable reasons of shared home co-habitants' perceived higher S\&P risks \textcolor{black}{\textbf{to their devices in shared home}}}
\label{tab:DeviceConcern}
{\begin{tabularx}{\linewidth}{Xccc}
\toprule
\textbf{S\&P Concern} & \textbf{$\chi^2$ Statistic} & \textbf{df} & \textbf{$p$-Value (2-sided)}\\
\midrule
(*) Staying in a privately rented house with multiple tenants & 14.217 & 3 & .003\\
(*) Not receiving relevant information from the landlord-owned and controlled devices & 13.935 & 6 & .030\\
(*) landlord-owned and controlled devices & 11.950 & 3 & .008\\
Routers owned by landlords and cohabitants & 17.446 & 8 & .026\\
(*) Access to the recording of the smart camera(s)/video doorbell(s) by landlord and other cohabitants & 170.600 & 18 & $<.001$\\
Shared WiFi passwords between occupants & 14.927 & 6 & .021\\
Shared WiFi passwords with visitors of co-occupants & 31.504 & 9 & $<.001$\\
Exposed WiFi password by the common placement of router & 152.474 & 9 & $<.001$\\
Being spied upon when devices are shared & 12.040 & 2 & .002\\
Sharing of smart hubs between co-occupants & 160.288 & 12 & $<.001$\\
Sharing smartphone with other co-habitants & 12.039 & 4 & .017\\
The need to hide the devices when there is a party/get-together organised by other co-habitants & 171.367 & 12 & $<.001$\\
S\&P is adversely affected by the negligence of co-occupants & 154.754 & 6 & $<.001$ \\
Malware affected devices of co-occupants might affect other devices at home & 159.407 & 12 & $<.001$\\
S\&P risks related to other co-habitants & 10.911 & 2 & .004\\
S\&P risks related to other co-habitants' visitors & 11.856 & 2 & .003\\
Devices being tampered with when not at home & 8.350 & 2 & .015\\
The WiFi network is not secure in a shared home & 5.535 & 21 & $<.001$\\
\bottomrule
\end{tabularx}}
\end{table}

Users we observed in our Reddit data were sometimes afraid of/concerned about their roommates as they could not convince them to adopt secure behaviours. We came across Reddit users expressing their concerns about their roommates who exhibited insecure behaviour by opening up random websites and installing software without checking their authenticity. Such posters were anxious and worried that they were unable to educate their roommate and that their roommates' devices would get infected by such insecure actions, and, consequently, they could get their own devices infected. At times, some posters feared that their co-habitants might hack their devices but were unsure how that might happen. They took to the Reddit platform to query about possible ways how they might be affected and likely devices that might be in use to gear such activities.

\subsubsection{\textbf{Cyber-Physical Threats and Concerns Thereof}:}

We observed that users in a shared home displayed concerns such as power imbalance between users, direct conflicts between users of different shared devices, and complex trust relationships amongst co-habitants of the home, which are somewhat similar and prevalent in traditional family homes~\citep{zeng2019HomeUser, Bernd2022MU}, although the trust level of participants is different. For example, the trust level between a resident and their co-habitants or their co-habitants' visitors is totally different from that between family members.

Our Reddit dataset includes several posts that discussed concerns about physical security. This echoes the findings of another paper~\citep{zeng2017end}, where physical security has also been recognised as the most common type of concern in a traditional family home setting. Our data show evidence of three different types of cyber-related threats described below: cyber threats, cyber-physical threats, and physical-cyber threats. For the latter two types of threats, we followed the \textit{Cyber Physical Threat} taxonomy proposed by~\cite {heartfield2018taxonomy}, who provided detailed examples of how security breaches in smart homes can affect both cyber and physical spaces.

\textit{Cyber threats}: These refer to types of potential S\&P attacks that might be happening in the cyber space~\cite{parikh2017cyber}. These include, for example, the fear of a shared home user from other co-habitants or their visitors' accessing their private information by accessing shared computing devices.

\textit{Cyber-physical threats}: These are threats where the physical safety of a shared home user is threatened by one or more smart devices that are affected by malicious cyber activities. We considered a \textit{cyber-physical} threat in line with the definition of `cyber-physical attack' by~\citet{loukas2015cyber} as a security threat from the cyber space that threatens to adversely affect the physical space. The affected people in the physical space are concerned about their physical privacy, safety, and well-being. For instance, in a Reddit post in our dataset, one shared home user complained about their roommate who supposedly violated the poster's privacy by hovering near their bedroom door. The poster enquired about how to install a motion detector to a smart bulb to automatically flash blind their roommate with bright light when they reach the bedroom door. 

\textit{Physical-cyber threats}: These are threats opposite to cyber-physical ones, where an attack performed in the physical space adversely affects the cyber space~\cite{loukas2015cyber}. Some posts in our Reddit data are examples of one sub-type of such threats called `Unauthorised actuation' reported by~\citet{heartfield2018taxonomy}, where the user's roommate kept disconnecting their security camera in the common living room, an unauthorised physical action by the roommate that led to concerns over the proper working of a smart home device.

Several examples (11/46, 23.9\%) touched upon the subject of privacy violation, where the poster's privacy was violated by the act of their co-habitants, but they were not sure of how to handle the situation (e.g., whether to contact the police and what are the grounds of such calls) or did not have the means to do so (e.g., lack of finance to access legal help). Other examples present evidence of privacy concerns and impact on domestic life, including `loss of control', `inconvenience' and `invasion of privacy' (terminology used by~\citet{heartfield2018taxonomy}).

\subsubsection{\textbf{S\&P Concerns and Threat Actors in a Shared Home}:}

Threat models in multi-user home security have been studied by many researchers~\cite{zeng2017end, he2018rethinking, Watson2020MU, slupska2019safe, slupska2021threat}. Threat modelling in smart homes generally espouses two main classes of adversaries, i.e., external parties who can cause damage to the system, data, and/or platforms, and internal adversaries, including household members who have physical access to the home and can pose a threat to all assets. Some researchers, such as~\citet{slupska2019safe}, focused on the extreme end of the concept of `insider threat' in smart homes while researching intimate partner abuse (IPV). At the same time, others~\cite{zeng2019HomeUser} discussed non-adversarial threats in a family home setting involving conflicts between different users. However, apart from~\citet{HE2021MU}, who developed a threat model on non-technical adversaries who have legitimate access to the home, no other prominent studies have looked into this area.

\begin{figure}[!htb]
\centering
\includegraphics[width=0.8\linewidth]{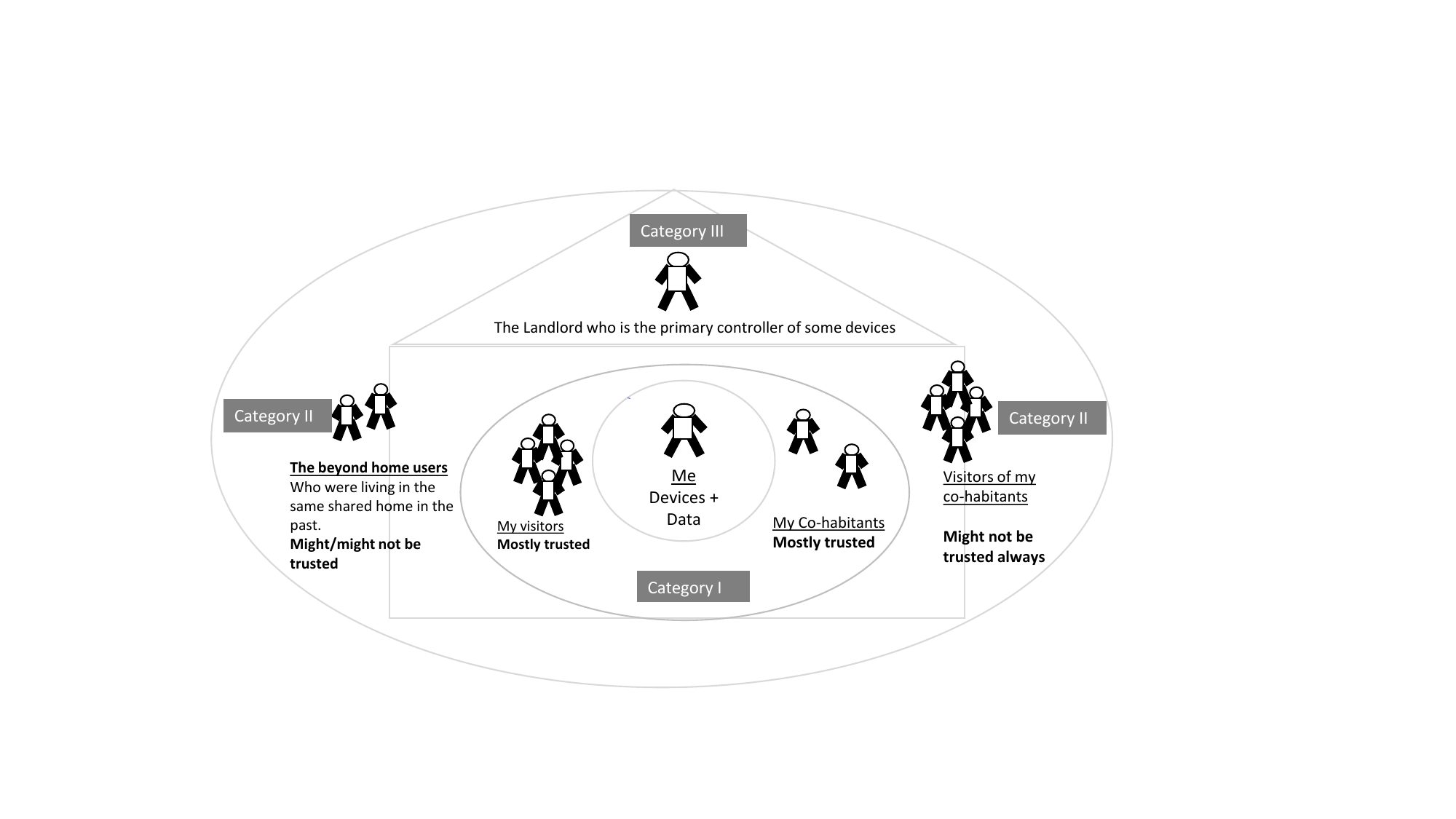}
\caption{Different threat actors in a shared home setting}
\label{fig:ThreatActors}
\end{figure}

Exploring both the survey data and the Reddit posts, we discovered different threat actors specific to shared home settings who might pose S\&P concerns for shared home users. Based on the data we collected and the information on related cases discussed in Section~\ref{Sec:RelatedWork}, we constructed a model representing possible threat actors which the shared home user might face internally under the umbrella of `insider threat.' The model is presented in Figure~\ref{fig:ThreatActors}. The model does not consider any outsider threat actors, e.g., hackers trying to sniff the house data, but focuses mainly on people who might be present in a shared home due to either being a resident or visiting the shared home or having a chance to visit the shared home legally/illegally due to their earlier connection with the shared home. We categorized the threat actors in the model into three categories, I, II, and III, according to the trust level that was revealed in our dataset and our analysis of it. As the scope of the threat actors broadens (this is represented with the grey ellipses in Figure~\ref{fig:ThreatActors}) to include the co-habitants and the shared home users' own visitors first, then the visitors of the co-habitants, the trust from these actors decreases, and correspondingly the security concern increases. We can see a reflection of this in our survey data where in answer to the question on whether users were concerned about S\&P risks, 5.3\% of the users showed concerns about their own visitors, 11.5\% were concerned about their co-occupants, and 16\% were concerned about the visitors of their co-occupants. Figure~\ref{fig:ThreatActors},  shows different threat actors in a shared home environment organized around a specific user in the shared home, who is shown in the middle of the diagram as `Me'. Each threat actor in the diagram is placed in a different category depending on to what extent the specific user trusts the threat actor. Other co-habitants and visitors of the specific user are placed in Category~I (the most trusted people by the specific user `Me'). Visitors of other co-habitants and the `beyond home users' are placed in Category~II (the group of people less trusted than Category~I people). The `beyond home users' are people who had access to the same shared home in the past as a former co-habitant or due to other reasons (e.g., a frequent visitor to the shared home in the past due to a close tie with a former co-habitant). The landlord is assigned a different category III owing to their special status in the shared home context.\footnote{Note that some landlords may rent one or more rooms in their home to others so they can also be co-habitants of the same shared home. For this paper, we did not consider such complicated scenarios.}

Some of our Reddit posts (7/46, 15.2\%) exhibit characteristics of what we have termed as the `beyond home' threat actors. These are people other than the landlord who do not live in the home but could visit and control one or more home devices in some way. For example, some posters expressed their concerns and frustration when their home devices were controlled and managed by their co-habitant's friends, whom the co-habitant trusted with the device management permission, without consulting everyone in the house. The `friend', in this case, was outside of the shared home but still controlled devices inside the home. In other cases, such posters were anxious about their security and privacy from their ex-roommates, who still had legal access to their shared home. Threats such as these cause concerns for people and still happen within the home boundary but are caused by actors beyond the home.

In summary, our observation of shared home privacy and security concerns evidenced that residents in a shared home environment face different threat actors due to the nature of their relationships with other co-habitants, which affect their trust level to a certain extent. The concepts of `insider threat' and `beyond home threat actors' in relation to shared homes are important subjects of further discussion and should be paid more attention to in future research. We also noticed that cyber-physical threats are a big concern for the residents of a shared home. Multi-user relationships emerging from the way devices are used, such as \textit{Primary-Primary}, \textit{ExternalPrimary-Secondary} and \textit{Primary-Bystander}, are some of the areas where we need to look at more closely to understand how such relationships impact the S\&P behaviours and actions of the residents as well as the landlords. Another intriguing point is that, although shared home users exhibit similar roles as bystanders and primary users, the contextual variations, in this case, can let people have different perspectives as bystanders. More specifically, in past studies~\cite{bernd2019BystanderDomestic, Markey2022RolesMatter}, bystanders considered are mostly visitors whose exposure to home devices as bystanders is accidental and infrequent, which is substantially different from exposure of shared home residents who have to face the situation on a daily basis. Research in this area would shed light on different factors that might influence or impact users' S\&P perspectives as a whole.

\section{Further Discussions, Limitations \& Future Work}
\label{Sec:Discussion}

Our focused investigation in this specific area of multi-user relationships in a shared home setting revealed a unique contextual disposition of user-to-user and user-to-device relationships regarding cyber security and privacy of the home. It uncovered some ignored or insufficiently studied aspects in past research on home S\&P, such as the landlords as \textbf{\textit{ExternalPrimary}} users, S\&P nuances of multiple primary users, `bystander' users beyond one-off or less frequent visitors, ex-inhabitants' ability and probability to access household data and/or the home, and consideration of various threat actors in multi-user settings. In addition, our research also led to a number of recommended future research directions, such as more research on specific demographic factors within the shared home population, more empirical studies broadening the scope of online sourced data our study used, and development of a fuller threat model for the multi-user environments. The above aspects and recommendations are discussed below in detail.

\subsection{Shared Home S\&P: Differences in User Perspectives}

Careful consideration of different user roles that we observed in our data demonstrated many similarities of user roles and relationships between traditional family homes and shared home users, such as primary, secondary users and bystanders. We highlighted the case of \textit{\textbf{multiple primary users}} in the shared home situation, which has been mentioned in some previous studies~\cite{Jang2017Multi-user} specifically focusing on houses where the primary/administrative permissions might be with more than one person. The main difference we noticed in the case of a shared home environment is that if the landlord, as a primary user, shares the administrative permissions of a home device such as a router with their tenants in a shared home, they are expected to share such permissions with each of the tenants by default, as everyone has an equal right to such permissions as a resident. In a traditional family home, these permissions might be shared only if the need arises to share, and there is no compulsion to share such permissions amongst all members. However, these types of shared administrative permissions in shared homes may give rise to a number of problems in terms of S\&P, such as who would be held responsible if the lack of security of one administrative user created an S\&P problem for all users and how a consensus could be decided in relation to the S\&P of a device managed by multiple primary users. For instance, an erroneous administrative setting in a smart camera by one careless or less skilled user might affect the privacy of every member of a shared home. As we noticed in our survey data, only 33\% of participants discussed any kind of S\&P configuration issues with their co-habitants.

This difference in the contextual balance of power in different types of multi-user homes merits further investigation in future research.

Another interesting point to mention here is the context in which \textbf{\textit{PassiveSecondary}} or \textit{bystanders} are verbalised in the existing literature. This user type is primarily looked into as non-household users, such as visitors, neighbours, and household workers~\cite{Cobb2021MU, bernd2019BystanderDomestic, tabassum2023exploring}, who have no or less access to the home devices in question. In a shared home, all co-habitants could be considered `\textit{PassiveSecondary}' when they do not own a device but are exposed to its use on a daily basis. Therefore, \textbf{the location of a bystander in a shared home is not necessarily external to the home.}

This leads to two relevant points that are of interest for future research: 1) the internality of the problem requires looking at \textbf{\textit{trust in a shared home and its significance in maintaining the social bond between co-habitants}}, and 2) the fact that shared home users are staying at home more on a daily basis justifies the consideration of the \textit{\textbf{topic of temporality}}.

The concept of trust in multi-user homes has been discussed by other researchers from the point of view of users' trust as providers of computing devices in a multi-user home~\cite{Meng2021MU, Ahmad_2020_TangibleMuser, Windl2022MU}, trust between the Airbnb host and guests~\cite{dey2020exploring, mare2020smart}, trust relationships between family members~\cite{matthews2016she, he2018rethinking, geeng2019Multi_User, Leitao2019IPA, al2021shared, koshy2021passenger}, and trust between the employer and employees in a home context~\cite{Bernd2022MU}. However, shared home users could have specific trust issues. For example, trusting other users with private resources when using the same WiFi, trust implications when device ownership/use of individual members could affect the daily activities of other users, i.e., keeping a camera to record the movements of other users, and trust between users when each one could possess primary user rights are some of the topics that need exploring further.

In considering trust between individuals and their co-habitants in a shared home, Altman's~\cite{altman1975environment} concept of `interpersonal boundary' is important. Altman discusses the dynamic negotiation process by the individuals to maintain their private space. This process is more challenging when we consider the use of smart devices as well at home. As Ahmad~\cite{Ahmad_2020_TangibleMuser} points out, the `software-controlled and display-limited' nature of smart devices creates a level of mismatch between the \textit{perceived} and \textit{actual} level of privacy for the users at home. So, considering from the `boundary regulation' perspective, we not only have to attend to the interpersonal relationship between individuals within the shared home but also address the trust and privacy expectations of the shared home users (or non-users in case of \textit{PassiveSecondary} residents) of the different computing devices present within that environment. The entanglement and complexities of these factors increase due to different types of users in a shared home environment, i.e. \textbf{multiple primary users}. Understanding the nuances involved in trust between the \textit{user -- user} and \textit{user -- smart device} in a shared home with multiple devices and their implications needs further research.

For the second point, which is the topic of temporality, discussions on \textit{bystanders} in the existing literature always have the flavour of a user who spends time with a certain device for a set temporary time period and gets exposed to the device. However, shared home residents are not temporary visitors, and they normally stay at home much longer than visitors. Therefore, they should be more concerned according to research of~\citet{yao2019privacy}, which showed that \textit{bystanders} are more concerned about their privacy when the `length of stay' is longer. Hence, further research is definitely needed for a better understanding of shared home users as different types of bystander users or what we termed as \textit{PassiveSecondary} users, which have not been studied in the past. This temporality issue is further convoluted by the ex-inhabitants' ability and probability to access household data. According to the English housing survey from 2021 -- 2022\footnote{\url{https://www.gov.uk/government/statistics/english-housing-survey-2021-to-2022-private-rented-sector/english-housing-survey-2021-to-2022-private-rented-sector\#housing-history-and-future-housing-aspirations}}, 88\% of the rental properties are rented as shorthold tenancy, i.e., 12 or 6 months. This is just evidence of the frequency with which the tenants in some of the regions might move. In these cases, there is always an increasing possibility of an S\&P breach from the ex-tenants, e.g., accessing the house wifi if the shared password is not changed or accessing the house physically if the smart lock password remains the same. Further research in this area is needed to understand and establish clear policies with regard to the changeover of tenants.

\subsection{S\&P Issues in Rental Homes}

The \textbf{\textit{landlord-tenant relationship}} highlighted in our study has not appeared in detail in past studies. Studies that looked into S\&P aspects of landlords and tenants focused on corporate landlords and broader technological facilitation on increasing rent revenues, such as management of the maintenance of the properties using smart apps or finding viable renters using \textit{``data-enabled disciplinary capabilities}''~\cite{nethercote2023platform}\footnote{\color{black}This is a method whereby the corporate landlords use the rental data at their disposal to screen their prospective renters.}. Researchers such as~\citet{hulse2014secure} and \citet{byrne2022secure}, who did look into rental security, explored the concept of rental tenure security and rental occupancy security on account of factors such as legislative, public policy, market force and cultural dimension. We did not notice any past studies that addressed S\&P aspects in the context of shared rental homes, considering the use of multiple computing devices by multiple users. We discussed the \textit{ExternalPrimary} user type in connection to the landlord, which might introduce a very different dimension to the existing multi-user relationships.

We also noted that, despite owning some devices in these premises, landlords hardly inform their tenants or sign any formal agreements with them concerning the use of home devices and the collection of related data (if any). An examination of the UK Government's landlord legal requirements~\cite{Gov202019LandlordRespon} revealed that landlords are not legally required to provide any such information to their tenants. We observed in our survey results that 30\% of the participants' data in the form of recording was collected by the landlord (for cases where a smart camera and/or video doorbell existed). The other important point that we would like to mention is that, although we investigated the landlord-tenant relationship in the context of non-family shared homes, a similar relationship is equally possible in the context of traditional family homes that are rented from a landlord.

Hence, we believe that the S\&P implications of the landlord-tenant relationship in both shared and traditional family homes need further research. This could include topics from the technical point of view, such as to what extent the landlord can exert their administrative power in configuring and setting the devices, legal obligations and rights the landlord has to store/use the tenants' data, and to what extent the tenants could protect their rights to privacy in terms of smart device data usage. 

\subsection{Insufficient research on smart home threat modelling}

We observed from our literature review that past research has not covered the different threat actors in multi-user homes, such as shared homes, focused in our study. \citet{HE2021MU} proposed a threat model focusing on non-expert adversaries in home settings, added on top of expert adversaries commonly considered in the past. However, their work focuses on contextual detection of access control in home IoT sensors. Our current study endeavours to consider all types of computing devices at home, including smart and non-smart devices, while building the threat model. Discussing their proposed threat model for smart home environments, \citet{zeng2017end} reflected on the fact that threat models often depend on the `sophistication' of the mental model of the users, and they stated that the interaction between the primary and other users could be more positive if they share the same threat model. \citet{huang_2020_amazon} observed how concerns about visitors and other users based on conjectures can lead to inaccurate or incorrect threat models. In her 2019 paper, \citet{slupska2019safe} raised concerns about the fact that how domestic technology abuse has been conspicuously absent from the general security threat models. They advocated for the same level of policies and protections for domestic abuse victims as happens at an organisational level and pointed out that such steps would only be possible when the `insider threats' model is considered seriously for intimate partner victims (IPVs) as an extreme type of adversarial settings in a home. Our work further advances our understanding of threat models for multi-user homes by focusing on shared home environments, a very specific area of multi-user homes that have not been studied before, and analysing the threat actors in such environments in detail. The results of our examination and previous studies both evidenced that threat modelling of multi-user home environments needs more attention from the research community as well as from smart home device designers and vendors.

One of the major contributions of our study is the identification of the \textit{insider threat actors} in a shared home environment. However, these threat actors are only part of the whole threat model, and our study examined only people who have access to computing devices at home. We acknowledge that there are some other threat actors our study did not consider, e.g., hackers who never stayed at a shared home but can still access some home devices illegally, vendors and other third parties who might also access some home devices in certain circumstances. These actors do not fall into the scope of our study as we primarily looked into users who are currently staying at a shared home, stayed previously, visited the house or have constant access to the home, such as the landlords. Our future avenue for research is to build a fuller threat model for the shared home environment and extend it to understand the threat model for the multi-user home environment. This would enable us to develop a more comprehensive threat outlook of the multi-user home and provide us with richer insights into multi-user S\&P perspectives.

\subsection{Limitations and Future Research Directions}

A majority of our survey data (64\%) was collected from UK participants, and hence, the data may lack the perception of people from \textcolor{black}{non-UK residents.} Prior research~\cite{Gupta2019SHCrossCultural} has evidenced how different cultural dimensions, such as masculinism and uncertainty avoidance, influence individual intentions to participate in a shared economy. Further research should incorporate participants from different geographical areas and culturally varied populations, which might reveal how cultural factors are important to understanding shared home users' S\&P behaviours. \textcolor{black}{Further studies focusing on more specific demographic factors within the shared home population may reveal interesting details on various aspects, e.g., whether the age or gender of cohabitants in a non-family household has any specific impact on their S\&P behaviours and/or concerns in general.}

\textcolor{black}{Additionally, our online data analysis specifically focused on Reddit. This choice has its limitations. Research has shown that different online platforms attract varying populations to their sites, leading to platform coverage error where the platform base is not aligned with the target population~\cite{sen2021total}. For Reddit, its population has been identified as primarily male and belongs to the younger generation\footnote{\url{https://www.socialchamp.io/blog/reddit-demographics/}}. Although we used relevant keywords to identify a sub-population on Reddit that had relevant discussions, we acknowledge that such a sub-population may be a biased representation of the whole target population of our study (all people living in a shared home environment). This limitation is, however, largely unavoidable because all online platforms have the same problem, and there are no obvious alternative channels for us to reach the target population directly. In our future work, we plan to consider other online platforms in order to cross-validate the Reddit-based results reported in this paper. We also plan to run some empirical studies such as surveys and interviews to overcome the problem of relying on online platform data and to identify other ways to reach out to more people in the target population.}

Due to the relatively small size of our dataset, our study might not be very representative of the perception of shared home users, but it raises important points on S\&P behaviours of rented shared home users, including the landlord-tenant relationship and possible threat actors inside a shared home. Additionally, it would be interesting to extend the participants to include multi-user views from traditional family settings as well to get an overall S\&P perception of multi-user residents in rented accommodations. Furthermore, the current work specifically focuses on privately rented homes, as more than 80\% of our participants lived in such homes. It would be interesting to explore specific communal living, such as student or staff accommodations and/or hostels, to understand whether the S\&P perspectives of these places differ from the current study as the control of devices and resources in these kinds of homes are even more centralised; the WiFi is shared with a comparatively larger group of people than a privately rented shared accommodation, which may increase S\&P vulnerabilities of shared home users.

\section{Conclusion}

Understanding the perspectives of shared home users is an important part of understanding the multi-user perspectives in a smart home. Using a survey and analysis of posts from the online forum Reddit, this study identified several relevant S\&P behaviours and concerns of residents of shared homes, such as the prevalence of multiple primary users and concerns due to cyber-physical threats. The research reflected on threat actors and analysed the `insider threats' in connection to a smart home, which are specific to a shared home setting while commenting on specific user roles, their contextual variations, and the unique S\&P role held by the landlords. The study suggested further research into insider threats specific to shared homes and exploring the varying roles of shared home users, considering the burgeoning interest in the area.

\bibliographystyle{ACM-Reference-Format}
\bibliography{main}


\begin{thebibliography}{93}


\ifx \showCODEN    \undefined \def \showCODEN     #1{\unskip}     \fi
\ifx \showDOI      \undefined \def \showDOI       #1{#1}\fi
\ifx \showISBNx    \undefined \def \showISBNx     #1{\unskip}     \fi
\ifx \showISBNxiii \undefined \def \showISBNxiii  #1{\unskip}     \fi
\ifx \showISSN     \undefined \def \showISSN      #1{\unskip}     \fi
\ifx \showLCCN     \undefined \def \showLCCN      #1{\unskip}     \fi
\ifx \shownote     \undefined \def \shownote      #1{#1}          \fi
\ifx \showarticletitle \undefined \def \showarticletitle #1{#1}   \fi
\ifx \showURL      \undefined \def \showURL       {\relax}        \fi
\providecommand\bibfield[2]{#2}
\providecommand\bibinfo[2]{#2}
\providecommand\natexlab[1]{#1}
\providecommand\showeprint[2][]{arXiv:#2}

\bibitem[Ahmad et~al\mbox{.}(2020)]%
        {Ahmad_2020_TangibleMuser}
\bibfield{author}{\bibinfo{person}{Imtiaz Ahmad}, \bibinfo{person}{Rosta Farzan}, \bibinfo{person}{Apu Kapadia}, {and} \bibinfo{person}{Adam~J. Lee}.} \bibinfo{year}{2020}\natexlab{}.
\newblock \showarticletitle{Tangible Privacy: Towards User-Centric Sensor Designs for Bystander Privacy}.
\newblock \bibinfo{journal}{\emph{Proceedings of the ACM on Human-Computer Interaction}} \bibinfo{volume}{4}, \bibinfo{number}{CSCW2}, Article \bibinfo{articleno}{116} (\bibinfo{year}{2020}), \bibinfo{numpages}{28}~pages.
\newblock
\urldef\tempurl%
\url{https://doi.org/10.1145/3415187}
\showDOI{\tempurl}


\bibitem[Al-Ameen et~al\mbox{.}(2021)]%
        {al2021shared}
\bibfield{author}{\bibinfo{person}{Mahdi~Nasrullah Al-Ameen}, \bibinfo{person}{Huzeyfe Kocabas}, \bibinfo{person}{Swapnil Nandy}, {and} \bibinfo{person}{Tanjina Tamanna}.} \bibinfo{year}{2021}\natexlab{}.
\newblock \showarticletitle{``We, three brothers have always known everything of each other''”: A Cross-cultural Study of Sharing Digital Devices and Online Accounts}.
\newblock \bibinfo{journal}{\emph{Proceedings on Privacy Enhancing Technologies}} \bibinfo{volume}{2021}, \bibinfo{number}{4} (\bibinfo{year}{2021}), \bibinfo{pages}{203--224}.
\newblock
\urldef\tempurl%
\url{https://doi.org/10.2478/popets-2021-0067}
\showDOI{\tempurl}


\bibitem[Alase(2017)]%
        {alase2017interpretative}
\bibfield{author}{\bibinfo{person}{Abayomi Alase}.} \bibinfo{year}{2017}\natexlab{}.
\newblock \showarticletitle{The Interpretative Phenomenological Analysis ({IPA}): A Guide to a Good Qualitative Research Approach}.
\newblock \bibinfo{journal}{\emph{International Journal of Education and Literacy Studies}} \bibinfo{volume}{5}, \bibinfo{number}{2} (\bibinfo{year}{2017}), \bibinfo{pages}{9--19}.
\newblock
\urldef\tempurl%
\url{https://doi.org/10.7575/aiac.ijels.v.5n.2p.9}
\showDOI{\tempurl}


\bibitem[Altman(1975)]%
        {altman1975environment}
\bibfield{author}{\bibinfo{person}{Irwin Altman}.} \bibinfo{year}{1975}\natexlab{}.
\newblock \showarticletitle{The environment and social behavior: privacy, personal space, territory, and crowding.}
\newblock  (\bibinfo{year}{1975}).
\newblock


\bibitem[Altman(1977)]%
        {altman1977privacy}
\bibfield{author}{\bibinfo{person}{Irwin Altman}.} \bibinfo{year}{1977}\natexlab{}.
\newblock \showarticletitle{Privacy regulation: Culturally universal or culturally specific?}
\newblock \bibinfo{journal}{\emph{Journal of Social Issues}} \bibinfo{volume}{33}, \bibinfo{number}{3} (\bibinfo{year}{1977}), \bibinfo{pages}{66--84}.
\newblock
\urldef\tempurl%
\url{https://doi.org/10.1111/j.1540-4560.1977.tb01883.x}
\showDOI{\tempurl}


\bibitem[Apthorpe et~al\mbox{.}(2022)]%
        {Apthorpe2022MU}
\bibfield{author}{\bibinfo{person}{Noah Apthorpe}, \bibinfo{person}{Pardis Emami-Naeini}, \bibinfo{person}{Arunesh Mathur}, \bibinfo{person}{Marshini Chetty}, {and} \bibinfo{person}{Nick Feamster}.} \bibinfo{year}{2022}\natexlab{}.
\newblock \showarticletitle{You, Me, and {IoT}: How Internet-connected Consumer Devices Affect Interpersonal Relationships}.
\newblock \bibinfo{journal}{\emph{ACM Transactions on Internet of Things}}  \bibinfo{volume}{3} (\bibinfo{year}{2022}).
\newblock
Issue 4.
\urldef\tempurl%
\url{https://doi.org/10.1145/3539737}
\showDOI{\tempurl}


\bibitem[Bernd et~al\mbox{.}(2022)]%
        {Bernd2022MU}
\bibfield{author}{\bibinfo{person}{Julia Bernd}, \bibinfo{person}{Ruba Abu-Salma}, \bibinfo{person}{Junghyun Choy}, {and} \bibinfo{person}{Alisa Frik}.} \bibinfo{year}{2022}\natexlab{}.
\newblock \showarticletitle{Balancing Power Dynamics in Smart Homes: Nannies' Perspectives on How Cameras Reflect and Affect Relationships Balancing Power Dynamics in Smart Homes: Nannies' Perspectives on How Cameras Reflect and Affect Relationships}. In \bibinfo{booktitle}{\emph{Proceedings of the 18th Symposium on Usable Privacy and Security}}. \bibinfo{publisher}{USENIX Association}.
\newblock
\urldef\tempurl%
\url{https://www.usenix.org/conference/soups2022/presentation/bernd}
\showURL{%
\tempurl}


\bibitem[Bernd et~al\mbox{.}(2020)]%
        {bernd2020bystanders}
\bibfield{author}{\bibinfo{person}{Julia Bernd}, \bibinfo{person}{Ruba Abu-Salma}, {and} \bibinfo{person}{Alisa Frik}.} \bibinfo{year}{2020}\natexlab{}.
\newblock \showarticletitle{Bystanders' Privacy: The Perspectives of Nannies on Smart Home Surveillance}. In \bibinfo{booktitle}{\emph{Proceedings of the 10th USENIX Workshop on Free and Open Communications on the Internet}}. \bibinfo{publisher}{USENIX Association}, \bibinfo{numpages}{14}~pages.
\newblock
\urldef\tempurl%
\url{https://www.usenix.org/conference/foci20/presentation/bernd}
\showURL{%
\tempurl}


\bibitem[Bernd et~al\mbox{.}(2019)]%
        {bernd2019BystanderDomestic}
\bibfield{author}{\bibinfo{person}{Julia Bernd}, \bibinfo{person}{Alisa Frik}, \bibinfo{person}{Maritza Johnson}, {and} \bibinfo{person}{Nathan Malkin}.} \bibinfo{year}{2019}\natexlab{}.
\newblock \showarticletitle{Smart Home Bystanders: Further Complexifying a Complex Context}. In \bibinfo{booktitle}{\emph{Proceedings of CI Symposium 2019}}. \bibinfo{publisher}{PrivaCI}, \bibinfo{numpages}{6}~pages.
\newblock
\urldef\tempurl%
\url{https://privaci.info/symposium2/papers_and_slides/Sub_Bernd_et_al_Bystanders_CI_2019.pdf}
\showURL{%
\tempurl}


\bibitem[Bleu(2023)]%
        {Bleu2023RedditStatistics}
\bibfield{author}{\bibinfo{person}{Nicola Bleu}.} \bibinfo{year}{2023}\natexlab{}.
\newblock \bibinfo{title}{23+ {Reddit} Statistics For 2023: Users, Revenue, And Growth}.
\newblock
\newblock
\urldef\tempurl%
\url{https://startupbonsai.com/reddit-statistics/}
\showURL{%
\tempurl}


\bibitem[Braun et~al\mbox{.}(2019)]%
        {braun2019thematic}
\bibfield{author}{\bibinfo{person}{Virginia Braun}, \bibinfo{person}{Victoria Clarke}, \bibinfo{person}{Nikki Hayfield}, {and} \bibinfo{person}{Gareth Terry}.} \bibinfo{year}{2019}\natexlab{}.
\newblock \showarticletitle{Thematic Analysis 48}.
\newblock \bibinfo{journal}{\emph{Handbook of research methods in health social sciences}} (\bibinfo{year}{2019}), \bibinfo{pages}{843--860}.
\newblock
\urldef\tempurl%
\url{https://doi.org/10.1007/978-981-10-5251-4_103}
\showDOI{\tempurl}


\bibitem[BT(2020)]%
        {BT2020ConnectedDevice}
\bibfield{author}{\bibinfo{person}{BT}.} \bibinfo{year}{2020}\natexlab{}.
\newblock \bibinfo{title}{83\% of {UK} consumers are planning to gift a connected device as families get set for {Christmas}, according to study by {BT} Full Fibre}.
\newblock
\newblock
\urldef\tempurl%
\url{https://newsroom.bt.com/83-of-uk-consumers-are-planning-to-gift-a-connected-device-as-families-get-set-for-christmas-according-to-study-by-bt-full-fibre/}
\showURL{%
\tempurl}


\bibitem[Byrne and McArdle(2022)]%
        {byrne2022secure}
\bibfield{author}{\bibinfo{person}{Michael Byrne} {and} \bibinfo{person}{Rachel McArdle}.} \bibinfo{year}{2022}\natexlab{}.
\newblock \showarticletitle{Secure occupancy, power and the landlord-tenant relation: a qualitative exploration of the {Irish} private rental sector}.
\newblock \bibinfo{journal}{\emph{Housing Studies}} \bibinfo{volume}{37}, \bibinfo{number}{1} (\bibinfo{year}{2022}), \bibinfo{pages}{124--142}.
\newblock
\urldef\tempurl%
\url{https://doi.org/10.1080/02673037.2020.1803801}
\showDOI{\tempurl}


\bibitem[Chen et~al\mbox{.}(2022)]%
        {chen2022sharing}
\bibfield{author}{\bibinfo{person}{Jiayi Chen}, \bibinfo{person}{Urs Hengartner}, {and} \bibinfo{person}{Hassan Khan}.} \bibinfo{year}{2022}\natexlab{}.
\newblock \showarticletitle{Sharing without Scaring: Enabling Smartphones to Become Aware of Temporary Sharing}. In \bibinfo{booktitle}{\emph{Proceedings of the 18th Symposium on Usable Privacy and Security}}. \bibinfo{publisher}{USENIX Association}, \bibinfo{pages}{671--685}.
\newblock
\urldef\tempurl%
\url{https://www.usenix.org/conference/soups2022/presentation/chen}
\showURL{%
\tempurl}


\bibitem[Choe et~al\mbox{.}(2012)]%
        {Choe2012SensorProxies}
\bibfield{author}{\bibinfo{person}{Eun~Kyoung Choe}, \bibinfo{person}{Sunny Consolvo}, \bibinfo{person}{Jaeyeon Jung}, \bibinfo{person}{Beverly Harrison}, \bibinfo{person}{Shwetak~N. Patel}, {and} \bibinfo{person}{Julie~A. Kientz}.} \bibinfo{year}{2012}\natexlab{}.
\newblock \showarticletitle{Investigating Receptiveness to Sensing and Inference in the Home Using Sensor Proxies}. In \bibinfo{booktitle}{\emph{Proceedings of the 2012 ACM Conference on Ubiquitous Computing}}. \bibinfo{publisher}{ACM}, \bibinfo{pages}{61--70}.
\newblock
\urldef\tempurl%
\url{https://doi.org/10.1145/2370216.2370226}
\showDOI{\tempurl}


\bibitem[Clark et~al\mbox{.}(2019)]%
        {clark2019rosters}
\bibfield{author}{\bibinfo{person}{Vicky Clark}, \bibinfo{person}{Keith Tuffin}, \bibinfo{person}{Natilene Bowker}, {and} \bibinfo{person}{Karen Frewin}.} \bibinfo{year}{2019}\natexlab{}.
\newblock \showarticletitle{Rosters: Freedom, responsibility, and co-operation in young adult shared households}.
\newblock \bibinfo{journal}{\emph{Australian Journal of Psychology}} \bibinfo{volume}{71}, \bibinfo{number}{3} (\bibinfo{year}{2019}), \bibinfo{pages}{232--240}.
\newblock
\urldef\tempurl%
\url{https://doi.org/10.1111/ajpy.12238}
\showDOI{\tempurl}


\bibitem[Cobb et~al\mbox{.}(2021a)]%
        {Cobb_2021_MultiUser}
\bibfield{author}{\bibinfo{person}{Camille Cobb}, \bibinfo{person}{Sruti Bhagavatula}, \bibinfo{person}{Kalil~Anderson Garrett}, \bibinfo{person}{Alison Hoffman}, \bibinfo{person}{Varun Rao}, {and} \bibinfo{person}{Lujo Bauer}.} \bibinfo{year}{2021}\natexlab{a}.
\newblock \showarticletitle{``{I} would have to evaluate their objections''": Privacy tensions between smart home device owners and incidental users}.
\newblock \bibinfo{journal}{\emph{Proceedings on Privacy Enhancing Technologies}} \bibinfo{volume}{2021}, \bibinfo{number}{4} (\bibinfo{year}{2021}), \bibinfo{pages}{54--75}.
\newblock
\urldef\tempurl%
\url{https://doi.org/10.2478/popets-2021-0060}
\showDOI{\tempurl}


\bibitem[Cobb et~al\mbox{.}(2021b)]%
        {Cobb2021MU}
\bibfield{author}{\bibinfo{person}{Camille Cobb}, \bibinfo{person}{Sruti Bhagavatula}, \bibinfo{person}{Kalil~Anderson Garrett}, \bibinfo{person}{Alison Hoffman}, \bibinfo{person}{Varun Rao}, {and} \bibinfo{person}{Lujo Bauer}.} \bibinfo{year}{2021}\natexlab{b}.
\newblock \showarticletitle{``I would have to evaluate their objections'': Privacy tensions between smart home device owners and incidental users}.
\newblock \bibinfo{journal}{\emph{Proceedings on Privacy Enhancing Technologies}} \bibinfo{volume}{2021}, \bibinfo{number}{4} (\bibinfo{year}{2021}), \bibinfo{pages}{54--75}.
\newblock
\urldef\tempurl%
\url{https://doi.org/10.2478/popets-2021-0060}
\showDOI{\tempurl}


\bibitem[Das et~al\mbox{.}(2019)]%
        {Das2019SocialInfluence}
\bibfield{author}{\bibinfo{person}{Sauvik Das}, \bibinfo{person}{Laura~A. Dabbish}, {and} \bibinfo{person}{Jason~I. Hong}.} \bibinfo{year}{2019}\natexlab{}.
\newblock \showarticletitle{A Typology of Perceived Triggers for End-User Security and Privacy Behaviors}. In \bibinfo{booktitle}{\emph{Proceedings of the 15th Symposium on Usable Privacy and Security}}. \bibinfo{publisher}{USENIX Association}, \bibinfo{pages}{97--115}.
\newblock
\urldef\tempurl%
\url{https://www.usenix.org/conference/soups2019/presentation/das}
\showURL{%
\tempurl}


\bibitem[Das et~al\mbox{.}(2014)]%
        {Das2014SocialInfluence}
\bibfield{author}{\bibinfo{person}{Sauvik Das}, \bibinfo{person}{Tiffany~Hyun-Jin Kim}, \bibinfo{person}{Laura~A. Dabbish}, {and} \bibinfo{person}{Jason~I. Hong}.} \bibinfo{year}{2014}\natexlab{}.
\newblock \showarticletitle{The Effect of Social Influence on Security Sensitivity}. In \bibinfo{booktitle}{\emph{Proceedings of the 10th Symposium On Usable Privacy and Security}}. \bibinfo{publisher}{USENIX Association}, \bibinfo{pages}{143--157}.
\newblock
\urldef\tempurl%
\url{https://www.usenix.org/conference/soups2014/proceedings/presentation/das}
\showURL{%
\tempurl}


\bibitem[{Department for Levelling Up, Housing and Communities and Ministry of Housing, Communities \& Local Government, UK Government}(2019)]%
        {Gov202019LandlordRespon}
\bibfield{author}{\bibinfo{person}{{Department for Levelling Up, Housing and Communities and Ministry of Housing, Communities \& Local Government, UK Government}}.} \bibinfo{year}{2019}\natexlab{}.
\newblock \bibinfo{title}{Guidance Landlord and tenant rights and responsibilities in the private rented sector}.
\newblock
\newblock
\urldef\tempurl%
\url{https://www.gov.uk/government/publications/landlord-and-tenant-rights-and-responsibilities-in-the-private-rented-sector/landlord-and-tenant-rights-and-responsibilities-in-the-private-rented-sector#landlords-rights-responsibilities-and-advice}
\showURL{%
\tempurl}


\bibitem[Dey et~al\mbox{.}(2020)]%
        {dey2020exploring}
\bibfield{author}{\bibinfo{person}{Rajib Dey}, \bibinfo{person}{Sayma Sultana}, \bibinfo{person}{Afsaneh Razi}, {and} \bibinfo{person}{Pamela~J Wisniewski}.} \bibinfo{year}{2020}\natexlab{}.
\newblock \showarticletitle{Exploring smart home device use by airbnb hosts}. In \bibinfo{booktitle}{\emph{Extended Abstracts of the 2020 CHI Conference on Human Factors in Computing Systems}}. \bibinfo{publisher}{ACM}, Article \bibinfo{articleno}{LBW082}, \bibinfo{numpages}{8}~pages.
\newblock
\urldef\tempurl%
\url{https://doi.org/10.1145/3334480.3382900}
\showDOI{\tempurl}


\bibitem[Druta and Ronald(2021)]%
        {druta2021SH}
\bibfield{author}{\bibinfo{person}{Oana Druta} {and} \bibinfo{person}{Richard Ronald}.} \bibinfo{year}{2021}\natexlab{}.
\newblock \showarticletitle{Living alone together in {Tokyo} share houses}.
\newblock \bibinfo{journal}{\emph{Social \& Cultural Geography}} \bibinfo{volume}{22}, \bibinfo{number}{9} (\bibinfo{year}{2021}), \bibinfo{pages}{1223--1240}.
\newblock
\urldef\tempurl%
\url{https://doi.org/10.1080/14649365.2020.1744704}
\showDOI{\tempurl}


\bibitem[Geeng and Roesner(2019)]%
        {geeng2019Multi_User}
\bibfield{author}{\bibinfo{person}{Christine Geeng} {and} \bibinfo{person}{Franziska Roesner}.} \bibinfo{year}{2019}\natexlab{}.
\newblock \showarticletitle{Who's In Control? Interactions In Multi-User Smart Homes}. In \bibinfo{booktitle}{\emph{Proceedings of the 2019 CHI Conference on Human Factors in Computing Systems}}. \bibinfo{publisher}{ACM}, \bibinfo{numpages}{13}~pages.
\newblock
\urldef\tempurl%
\url{https://doi.org/10.1145/3290605.3300498}
\showDOI{\tempurl}


\bibitem[{Greater London Authority}(2022)]%
        {London2022HMO}
\bibfield{author}{\bibinfo{person}{{Greater London Authority}}.} \bibinfo{year}{2022}\natexlab{}.
\newblock \bibinfo{title}{Housing in {London} 2022}.
\newblock
\newblock
\urldef\tempurl%
\url{https://data.london.gov.uk/housing/housing-in-london/}
\showURL{%
\tempurl}


\bibitem[Gupta et~al\mbox{.}(2019)]%
        {Gupta2019SHCrossCultural}
\bibfield{author}{\bibinfo{person}{Manjul Gupta}, \bibinfo{person}{Pouyan Esmaeilzadeh}, \bibinfo{person}{Irem Uz}, {and} \bibinfo{person}{Vanesa~M. Tennant}.} \bibinfo{year}{2019}\natexlab{}.
\newblock \showarticletitle{The effects of national cultural values on individuals' intention to participate in peer-to-peer sharing economy}.
\newblock \bibinfo{journal}{\emph{Journal of Business Research}}  \bibinfo{volume}{97} (\bibinfo{year}{2019}), \bibinfo{pages}{20--29}.
\newblock
\urldef\tempurl%
\url{https://doi.org/10.1016/j.jbusres.2018.12.018}
\showDOI{\tempurl}


\bibitem[He et~al\mbox{.}(2018)]%
        {he2018rethinking}
\bibfield{author}{\bibinfo{person}{Weijia He}, \bibinfo{person}{Maximilian Golla}, \bibinfo{person}{Roshni Padhi}, \bibinfo{person}{Jordan Ofek}, \bibinfo{person}{Markus D{\"u}rmuth}, \bibinfo{person}{Earlence Fernandes}, {and} \bibinfo{person}{Blase Ur}.} \bibinfo{year}{2018}\natexlab{}.
\newblock \showarticletitle{Rethinking Access Control and Authentication for the Home {Internet of Things} ({IoT})}. In \bibinfo{booktitle}{\emph{Proceedings of the 27th USENIX Security Symposium}}. \bibinfo{publisher}{USENIX Assocaition}, \bibinfo{pages}{255--272}.
\newblock
\urldef\tempurl%
\url{https://www.usenix.org/conference/usenixsecurity18/presentation/he}
\showURL{%
\tempurl}


\bibitem[He et~al\mbox{.}(2021)]%
        {HE2021MU}
\bibfield{author}{\bibinfo{person}{Weijia He}, \bibinfo{person}{Valerie Zhao}, \bibinfo{person}{Olivia Morkved}, \bibinfo{person}{Sabeeka Siddiqui}, \bibinfo{person}{Earlence Fernandes}, \bibinfo{person}{Josiah Hester}, {and} \bibinfo{person}{Blase Ur}.} \bibinfo{year}{2021}\natexlab{}.
\newblock \showarticletitle{{SoK}: Context Sensing for Access Control in the Adversarial Home {IoT}}. In \bibinfo{booktitle}{\emph{Proceedings of the 2021 IEEE European Symposium on Security and Privacy}}. \bibinfo{publisher}{IEEE}, \bibinfo{pages}{37--53}.
\newblock
\urldef\tempurl%
\url{https://doi.org/10.1109/EUROSP51992.2021.00014}
\showDOI{\tempurl}


\bibitem[Heartfield et~al\mbox{.}(2018)]%
        {heartfield2018taxonomy}
\bibfield{author}{\bibinfo{person}{Ryan Heartfield}, \bibinfo{person}{George Loukas}, \bibinfo{person}{Sanja Budimir}, \bibinfo{person}{Anatolij Bezemskij}, \bibinfo{person}{Johnny~R.J. Fontaine}, \bibinfo{person}{Avgoustinos Filippoupolitis}, {and} \bibinfo{person}{Etienne Roesch}.} \bibinfo{year}{2018}\natexlab{}.
\newblock \showarticletitle{A taxonomy of cyber-physical threats and impact in the smart home}.
\newblock \bibinfo{journal}{\emph{Computers \& Security}}  \bibinfo{volume}{78} (\bibinfo{year}{2018}), \bibinfo{pages}{398--428}.
\newblock
\urldef\tempurl%
\url{https://doi.org/10.1016/j.cose.2018.07.011}
\showDOI{\tempurl}


\bibitem[Hirsch and Silverstone(2003)]%
        {hirsch2003information}
\bibfield{author}{\bibinfo{person}{Eric Hirsch} {and} \bibinfo{person}{Roger Silverstone}.} \bibinfo{year}{2003}\natexlab{}.
\newblock \showarticletitle{Information and communication technologies and the moral economy of the household}.
\newblock In \bibinfo{booktitle}{\emph{Consuming Technologies}}. \bibinfo{publisher}{Routledge}, \bibinfo{pages}{25--40}.
\newblock


\bibitem[Huang et~al\mbox{.}(2020)]%
        {huang_2020_amazon}
\bibfield{author}{\bibinfo{person}{Yue Huang}, \bibinfo{person}{Borke Obada-Obieh}, {and} \bibinfo{person}{Konstantin~(Kosta) Beznosov}.} \bibinfo{year}{2020}\natexlab{}.
\newblock \showarticletitle{{Amazon} vs.\ My Brother: How Users of Shared Smart Speakers Perceive and Cope with Privacy Risks}. In \bibinfo{booktitle}{\emph{Proceedings of the 2020 CHI Conference on Human Factors in Computing Systems}}. \bibinfo{publisher}{ACM}, Article \bibinfo{articleno}{402}, \bibinfo{numpages}{13}~pages.
\newblock
\urldef\tempurl%
\url{https://doi.org/10.1145/3313831.3376529}
\showDOI{\tempurl}


\bibitem[Hulse and Milligan(2014)]%
        {hulse2014secure}
\bibfield{author}{\bibinfo{person}{Kath Hulse} {and} \bibinfo{person}{Vivienne Milligan}.} \bibinfo{year}{2014}\natexlab{}.
\newblock \showarticletitle{Secure occupancy: A new framework for analysing security in rental housing}.
\newblock \bibinfo{journal}{\emph{Housing Studies}} \bibinfo{volume}{29}, \bibinfo{number}{5} (\bibinfo{year}{2014}), \bibinfo{pages}{638--656}.
\newblock
\urldef\tempurl%
\url{https://doi.org/10.1080/02673037.2013.873116}
\showDOI{\tempurl}


\bibitem[Jacobs et~al\mbox{.}(2016)]%
        {jacobs2016caring}
\bibfield{author}{\bibinfo{person}{Maia Jacobs}, \bibinfo{person}{Henriette Cramer}, {and} \bibinfo{person}{Louise Barkhuus}.} \bibinfo{year}{2016}\natexlab{}.
\newblock \showarticletitle{Caring About Sharing: Couples' Practices in Single User Device Access}. In \bibinfo{booktitle}{\emph{Proceedings of the 2016 ACM International Conference on Supporting Group Work}}. \bibinfo{publisher}{ACM}, \bibinfo{pages}{235--243}.
\newblock
\urldef\tempurl%
\url{https://doi.org/10.1145/2957276.2957296}
\showDOI{\tempurl}


\bibitem[Jamnik and Lane(2019)]%
        {jamnik2019use}
\bibfield{author}{\bibinfo{person}{Matthew~R. Jamnik} {and} \bibinfo{person}{David~J. Lane}.} \bibinfo{year}{2019}\natexlab{}.
\newblock \showarticletitle{The use of {Reddit} as an inexpensive source for high-quality data}.
\newblock \bibinfo{journal}{\emph{Practical Assessment, Research, and Evaluation}} \bibinfo{volume}{22}, \bibinfo{number}{1}, Article \bibinfo{articleno}{5} (\bibinfo{year}{2019}), \bibinfo{numpages}{10}~pages.
\newblock
\urldef\tempurl%
\url{https://doi.org/10.7275/j18t-c009}
\showDOI{\tempurl}


\bibitem[Jang et~al\mbox{.}(2017)]%
        {Jang2017Multi-user}
\bibfield{author}{\bibinfo{person}{William Jang}, \bibinfo{person}{Adil Chhabra}, {and} \bibinfo{person}{Aarathi Prasad}.} \bibinfo{year}{2017}\natexlab{}.
\newblock \showarticletitle{Enabling Multi-User Controls in Smart Home Devices}. In \bibinfo{booktitle}{\emph{Proceedings of the 2017 Workshop on Internet of Things Security and Privacy}}. \bibinfo{publisher}{ACM}, \bibinfo{pages}{49--54}.
\newblock
\urldef\tempurl%
\url{https://doi.org/10.1145/3139937.3139941}
\showDOI{\tempurl}


\bibitem[Kanchi and Karlapalem(2021)]%
        {Kanchi2021MU}
\bibfield{author}{\bibinfo{person}{Shravya Kanchi} {and} \bibinfo{person}{Kamalakar Karlapalem}.} \bibinfo{year}{2021}\natexlab{}.
\newblock \showarticletitle{A Multi Perspective Access Control in a Smart Home}. In \bibinfo{booktitle}{\emph{Proceedings of the 11th ACM Conference on Data and Application Security and Privacy}}. \bibinfo{publisher}{ACM}, \bibinfo{pages}{321--323}.
\newblock
\urldef\tempurl%
\url{https://doi.org/10.1145/3422337.3450324}
\showDOI{\tempurl}


\bibitem[Komen(2016)]%
        {komen2016MobileSharing}
\bibfield{author}{\bibinfo{person}{Leah Komen}.} \bibinfo{year}{2016}\natexlab{}.
\newblock \showarticletitle{``Here you can use it'': Understanding mobile phone sharing and the concerns it elicits in rural {Kenya}}.
\newblock \bibinfo{journal}{\emph{for(e)dialogue}} \bibinfo{volume}{1}, \bibinfo{number}{1} (\bibinfo{year}{2016}), \bibinfo{pages}{52--65}.
\newblock
\urldef\tempurl%
\url{https://doi.org/10.29311/for(e)dialogue.v1i1.532}
\showDOI{\tempurl}


\bibitem[Koshy et~al\mbox{.}(2021a)]%
        {Koshey2021MU}
\bibfield{author}{\bibinfo{person}{Vinay Koshy}, \bibinfo{person}{Joon Sung~Sung Park}, \bibinfo{person}{Ti-Chung Cheng}, {and} \bibinfo{person}{Karrie Karahalios}.} \bibinfo{year}{2021}\natexlab{a}.
\newblock \showarticletitle{``We Just Use What They Give Us'': Understanding Passenger User Perspectives in Smart Homes}. In \bibinfo{booktitle}{\emph{Proceedings of the 2021 CHI Conference on Human Factors in Computing Systems}}. \bibinfo{publisher}{ACM}, Article \bibinfo{articleno}{41}, \bibinfo{numpages}{14}~pages.
\newblock
\urldef\tempurl%
\url{https://doi.org/10.1145/3411764.3445598}
\showDOI{\tempurl}


\bibitem[Koshy et~al\mbox{.}(2021b)]%
        {koshy2021passenger}
\bibfield{author}{\bibinfo{person}{Vinay Koshy}, \bibinfo{person}{Joon Sung~Sung Park}, \bibinfo{person}{Ti-Chung Cheng}, {and} \bibinfo{person}{Karrie Karahalios}.} \bibinfo{year}{2021}\natexlab{b}.
\newblock \showarticletitle{``We Just Use What They Give Us'': Understanding Passenger User Perspectives in Smart Homes}. In \bibinfo{booktitle}{\emph{Proceedings of the 2021 CHI Conference on Human Factors in Computing Systems}}. \bibinfo{publisher}{ACM}, Article \bibinfo{articleno}{41}, \bibinfo{numpages}{14}~pages.
\newblock
\urldef\tempurl%
\url{https://doi.org/10.1145/3411764.3445598}
\showDOI{\tempurl}


\bibitem[Kramer(2020)]%
        {PewResearchCenter2020home}
\bibfield{author}{\bibinfo{person}{Stephanie Kramer}.} \bibinfo{year}{2020}\natexlab{}.
\newblock \bibinfo{title}{With billions confined to their homes worldwide, which living arrangements are most common?}
\newblock
\newblock
\urldef\tempurl%
\url{https://www.pewresearch.org/short-reads/2020/03/31/with-billions-confined-to-their-homes-worldwide-which-living-arrangements-are-most-common/}
\showURL{%
\tempurl}


\bibitem[Lau et~al\mbox{.}(2018)]%
        {lau2018alexa}
\bibfield{author}{\bibinfo{person}{Josephine Lau}, \bibinfo{person}{Benjamin Zimmerman}, {and} \bibinfo{person}{Florian Schaub}.} \bibinfo{year}{2018}\natexlab{}.
\newblock \showarticletitle{{Alexa}, Are You Listening?: Privacy Perceptions, Concerns and Privacy-seeking Behaviors with Smart Speakers}.
\newblock \bibinfo{journal}{\emph{Proceedings of the ACM on Human-Computer Interaction}} \bibinfo{volume}{2}, \bibinfo{number}{CSCW}, Article \bibinfo{articleno}{102} (\bibinfo{year}{2018}), \bibinfo{numpages}{31}~pages.
\newblock
\urldef\tempurl%
\url{https://doi.org/10.1145/3274371}
\showDOI{\tempurl}


\bibitem[Leit\~{a}o(2019)]%
        {Leitao2019IPA}
\bibfield{author}{\bibinfo{person}{Roxanne Leit\~{a}o}.} \bibinfo{year}{2019}\natexlab{}.
\newblock \showarticletitle{Anticipating Smart Home Security and Privacy Threats with Survivors of Intimate Partner Abuse}. In \bibinfo{booktitle}{\emph{Proceedings of the 2019 ACM Designing Interactive Systems Conference}}. \bibinfo{publisher}{ACM}, \bibinfo{pages}{527--539}.
\newblock
\urldef\tempurl%
\url{https://doi.org/10.1145/3322276.3322366}
\showDOI{\tempurl}


\bibitem[Lin and Parkin(2020)]%
        {lin2020transferability}
\bibfield{author}{\bibinfo{person}{Vanessa~Z. Lin} {and} \bibinfo{person}{Simon Parkin}.} \bibinfo{year}{2020}\natexlab{}.
\newblock \showarticletitle{Transferability of privacy-related behaviours to shared smart home assistant devices}. In \bibinfo{booktitle}{\emph{Proceedings of the 2020 7th International Conference on Internet of Things: Systems, Management and Security}}. \bibinfo{publisher}{IEEE}, \bibinfo{numpages}{8}~pages.
\newblock
\urldef\tempurl%
\url{https://doi.org/10.1109/IOTSMS52051.2020.9340199}
\showDOI{\tempurl}


\bibitem[Loukas(2015)]%
        {loukas2015cyber}
\bibfield{author}{\bibinfo{person}{George Loukas}.} \bibinfo{year}{2015}\natexlab{}.
\newblock \bibinfo{booktitle}{\emph{Cyber-Physical Attacks: A Growing Invisible Threat}}.
\newblock \bibinfo{publisher}{Butterworth-Heinemann}.
\newblock


\bibitem[Maalsen(2020)]%
        {Maalsen2020SharedHousing}
\bibfield{author}{\bibinfo{person}{Sophia Maalsen}.} \bibinfo{year}{2020}\natexlab{}.
\newblock \showarticletitle{'Generation Share': digitalized geographies of shared housing}.
\newblock \bibinfo{journal}{\emph{Social \& Cultural Geography}} \bibinfo{volume}{21}, \bibinfo{number}{1} (\bibinfo{year}{2020}), \bibinfo{pages}{105--113}.
\newblock
\urldef\tempurl%
\url{https://doi.org/10.1080/14649365.2018.1466355}
\showDOI{\tempurl}


\bibitem[Maalsen(2023)]%
        {maalsen2023SH}
\bibfield{author}{\bibinfo{person}{Sophia Maalsen}.} \bibinfo{year}{2023}\natexlab{}.
\newblock \showarticletitle{`We're the cheap smart home': the actually existing smart home as rented and shared}.
\newblock \bibinfo{journal}{\emph{Social \& Cultural Geography}} \bibinfo{volume}{24}, \bibinfo{number}{8} (\bibinfo{year}{2023}), \bibinfo{pages}{1383--1402}.
\newblock
\urldef\tempurl%
\url{https://doi.org/10.1080/14649365.2022.2065693}
\showDOI{\tempurl}


\bibitem[Malkin et~al\mbox{.}(2019)]%
        {malkin2019privacy}
\bibfield{author}{\bibinfo{person}{Nathan Malkin}, \bibinfo{person}{Joe Deatrick}, \bibinfo{person}{Allen Tong}, \bibinfo{person}{Primal Wijesekera}, \bibinfo{person}{Serge Egelman}, {and} \bibinfo{person}{David Wagner}.} \bibinfo{year}{2019}\natexlab{}.
\newblock \showarticletitle{Privacy Attitudes of Smart Speaker Users}.
\newblock \bibinfo{journal}{\emph{Proceedings on Privacy Enhancing Technologies}} \bibinfo{volume}{2019}, \bibinfo{number}{4} (\bibinfo{year}{2019}), \bibinfo{pages}{250--271}.
\newblock
\urldef\tempurl%
\url{https://doi.org/10.2478/popets-2019-0068}
\showDOI{\tempurl}


\bibitem[Mare et~al\mbox{.}(2020)]%
        {mare2020smart}
\bibfield{author}{\bibinfo{person}{Shrirang Mare}, \bibinfo{person}{Franziska Roesner}, {and} \bibinfo{person}{Tadayoshi Kohno}.} \bibinfo{year}{2020}\natexlab{}.
\newblock \showarticletitle{Smart Devices in {Airbnbs}: Considering Privacy and Security for both Guests and Hosts}.
\newblock \bibinfo{journal}{\emph{Proceedings on Privacy Enhancing Technologies}} \bibinfo{volume}{2000}, \bibinfo{number}{2} (\bibinfo{year}{2020}), \bibinfo{pages}{436--458}.
\newblock
\urldef\tempurl%
\url{https://doi.org/10.2478/popets-2020-0035}
\showDOI{\tempurl}


\bibitem[Marky et~al\mbox{.}(2022a)]%
        {marky2022PrivacyTea}
\bibfield{author}{\bibinfo{person}{Karola Marky}, \bibinfo{person}{Paul Gerber}, \bibinfo{person}{Michelle~Gabriela Pelzer}, \bibinfo{person}{Mohamed Khamis}, {and} \bibinfo{person}{Max M{\"u}hlh{\"a}user}.} \bibinfo{year}{2022}\natexlab{a}.
\newblock \showarticletitle{``You offer privacy like you offer tea''”: Investigating Mechanisms for Improving Guest Privacy in {IoT}-Equipped Households}.
\newblock \bibinfo{journal}{\emph{Proceedings on Privacy Enhancing Technologies}} \bibinfo{volume}{2022}, \bibinfo{number}{4} (\bibinfo{year}{2022}), \bibinfo{pages}{400--420}.
\newblock
\urldef\tempurl%
\url{https://doi.org/10.56553/popets-2022-0115}
\showDOI{\tempurl}


\bibitem[Marky et~al\mbox{.}(2020a)]%
        {Markey_2020_YouJust_MultiUser}
\bibfield{author}{\bibinfo{person}{Karola Marky}, \bibinfo{person}{Sarah Prange}, \bibinfo{person}{Florian Krell}, \bibinfo{person}{Max M\"{u}hlh\"{a}user}, {and} \bibinfo{person}{Florian Alt}.} \bibinfo{year}{2020}\natexlab{a}.
\newblock \showarticletitle{``You Just Can't Know about Everything'': Privacy Perceptions of Smart Home Visitors}. In \bibinfo{booktitle}{\emph{Proceedings of the 19th International Conference on Mobile and Ubiquitous Multimedia}}. \bibinfo{publisher}{ACM}, \bibinfo{pages}{83--95}.
\newblock
\urldef\tempurl%
\url{https://doi.org/10.1145/3428361.3428464}
\showDOI{\tempurl}


\bibitem[Marky et~al\mbox{.}(2021)]%
        {marky2021roles}
\bibfield{author}{\bibinfo{person}{Karola Marky}, \bibinfo{person}{Sarah Prange}, \bibinfo{person}{Max M{\"u}hlh{\"a}user}, {and} \bibinfo{person}{Florian Alt}.} \bibinfo{year}{2021}\natexlab{}.
\newblock \showarticletitle{Roles matter! Understanding differences in the privacy mental models of smart home visitors and residents}. In \bibinfo{booktitle}{\emph{Proceedings of the 20th International Conference on Mobile and Ubiquitous Multimedia}}. \bibinfo{publisher}{ACM}, \bibinfo{pages}{108--122}.
\newblock
\urldef\tempurl%
\url{https://doi.org/10.1145/3490632.3490664}
\showDOI{\tempurl}


\bibitem[Marky et~al\mbox{.}(2022b)]%
        {Markey2022RolesMatter}
\bibfield{author}{\bibinfo{person}{Karola Marky}, \bibinfo{person}{Sarah Prange}, \bibinfo{person}{Max M\"{u}hlh\"{a}user}, {and} \bibinfo{person}{Florian Alt}.} \bibinfo{year}{2022}\natexlab{b}.
\newblock \showarticletitle{Roles Matter! Understanding Differences in the Privacy Mental Models of Smart Home Visitors and Residents}. In \bibinfo{booktitle}{\emph{Proceedings of the 20th International Conference on Mobile and Ubiquitous Multimedia}}. \bibinfo{publisher}{ACM}, \bibinfo{pages}{108–122}.
\newblock
\urldef\tempurl%
\url{https://doi.org/10.1145/3490632.3490664}
\showDOI{\tempurl}


\bibitem[Marky et~al\mbox{.}(2020b)]%
        {Markey_2020_Idont_MultiUser}
\bibfield{author}{\bibinfo{person}{Karola Marky}, \bibinfo{person}{Alexandra Voit}, \bibinfo{person}{Alina St\"{o}ver}, \bibinfo{person}{Kai Kunze}, \bibinfo{person}{Svenja Schr\"{o}der}, {and} \bibinfo{person}{Max M\"{u}hlh\"{a}user}.} \bibinfo{year}{2020}\natexlab{b}.
\newblock \showarticletitle{``{I} Don't Know How to Protect Myself'': Understanding Privacy Perceptions Resulting from the Presence of Bystanders in Smart Environments}. In \bibinfo{booktitle}{\emph{Proceedings of the 11th Nordic Conference on Human-Computer Interaction: Shaping Experiences, Shaping Society}}. \bibinfo{publisher}{ACM}, Article \bibinfo{articleno}{4}, \bibinfo{numpages}{11}~pages.
\newblock
\urldef\tempurl%
\url{https://doi.org/10.1145/3419249.3420164}
\showDOI{\tempurl}


\bibitem[Marky et~al\mbox{.}(2020c)]%
        {Markey2020AllinOne_Behavior}
\bibfield{author}{\bibinfo{person}{Karola Marky}, \bibinfo{person}{Verena Zimmermann}, \bibinfo{person}{Alina St\"{o}ver}, \bibinfo{person}{Philipp Hoffmann}, \bibinfo{person}{Kai Kunze}, {and} \bibinfo{person}{Max M\"{u}hlh\"{a}user}.} \bibinfo{year}{2020}\natexlab{c}.
\newblock \showarticletitle{All in One! User Perceptions on Centralized {IoT} Privacy Settings}. In \bibinfo{booktitle}{\emph{Extended Abstracts of the 2020 CHI Conference on Human Factors in Computing Systems}}. \bibinfo{publisher}{ACM}, Article \bibinfo{articleno}{LBW071}, \bibinfo{numpages}{8}~pages.
\newblock
\urldef\tempurl%
\url{https://doi.org/10.1145/3334480.3383016}
\showDOI{\tempurl}


\bibitem[Matthews et~al\mbox{.}(2016)]%
        {matthews2016she}
\bibfield{author}{\bibinfo{person}{Tara Matthews}, \bibinfo{person}{Kerwell Liao}, \bibinfo{person}{Anna Turner}, \bibinfo{person}{Marianne Berkovich}, \bibinfo{person}{Robert Reeder}, {and} \bibinfo{person}{Sunny Consolvo}.} \bibinfo{year}{2016}\natexlab{}.
\newblock \showarticletitle{``She'll just grab any device that's closer'': A Study of Everyday Device \& Account Sharing in Households}. In \bibinfo{booktitle}{\emph{Proceedings of the 2016 CHI Conference on Human Factors in Computing Systems}}. \bibinfo{publisher}{ACM}, \bibinfo{pages}{5921--5932}.
\newblock
\urldef\tempurl%
\url{https://doi.org/10.1145/2858036.2858051}
\showDOI{\tempurl}


\bibitem[McKay and Miller(2021)]%
        {MackayMiller2021AbuseVictim}
\bibfield{author}{\bibinfo{person}{Dana McKay} {and} \bibinfo{person}{Charlynn Miller}.} \bibinfo{year}{2021}\natexlab{}.
\newblock \showarticletitle{Standing in the Way of Control: A Call to Action to Prevent Abuse through Better Design of Smart Technologies}. In \bibinfo{booktitle}{\emph{Proceedings of the 2021 CHI Conference on Human Factors in Computing Systems}}. \bibinfo{publisher}{ACM}, Article \bibinfo{articleno}{332}, \bibinfo{numpages}{14}~pages.
\newblock
\urldef\tempurl%
\url{https://doi.org/10.1145/3411764.3445114}
\showDOI{\tempurl}


\bibitem[Meng(2021)]%
        {Meng2021MU}
\bibfield{author}{\bibinfo{person}{Nicole Meng}.} \bibinfo{year}{2021}\natexlab{}.
\newblock \showarticletitle{Owning and Sharing: Privacy Perceptions of Smart Speaker Users}.
\newblock \bibinfo{journal}{\emph{Proceedings of the ACM on Human-Computer Interaction}} \bibinfo{volume}{5}, \bibinfo{number}{CSCW1}, Article \bibinfo{articleno}{45} (\bibinfo{year}{2021}), \bibinfo{numpages}{29}~pages.
\newblock
\urldef\tempurl%
\url{https://doi.org/10.1145/3449119}
\showDOI{\tempurl}


\bibitem[Moh et~al\mbox{.}(2022)]%
        {moh2022characterizing}
\bibfield{author}{\bibinfo{person}{Phoebe Moh}, \bibinfo{person}{Pubali Datta}, \bibinfo{person}{Noel Warford}, \bibinfo{person}{Adam Bates}, \bibinfo{person}{Nathan Malkin}, {and} \bibinfo{person}{Michelle~L. Mazurek}.} \bibinfo{year}{2022}\natexlab{}.
\newblock \showarticletitle{Characterizing Everyday Misuse of Smart Home Devices}. In \bibinfo{booktitle}{\emph{Proceedings of the 2023 IEEE Symposium on Security and Privacy}}. \bibinfo{publisher}{IEEE}, \bibinfo{pages}{1558--1572}.
\newblock
\urldef\tempurl%
\url{https://doi.org/10.1109/SP46215.2023.00089}
\showDOI{\tempurl}


\bibitem[Mortan(2022)]%
        {BBC2022SharedHome}
\bibfield{author}{\bibinfo{person}{Becky Mortan}.} \bibinfo{year}{2022}\natexlab{}.
\newblock \bibinfo{title}{Over-50s turn to house-shares to beat rising rents}.
\newblock
\newblock
\urldef\tempurl%
\url{https://www.bbc.co.uk/news/business-62344571}
\showURL{%
\tempurl}


\bibitem[Nethercote(2019)]%
        {nethercote2019caring}
\bibfield{author}{\bibinfo{person}{Megan Nethercote}.} \bibinfo{year}{2019}\natexlab{}.
\newblock \showarticletitle{Caring households: The social ties that house}.
\newblock \bibinfo{journal}{\emph{Housing, Theory and Society}} \bibinfo{volume}{36}, \bibinfo{number}{3} (\bibinfo{year}{2019}), \bibinfo{pages}{257--273}.
\newblock
\urldef\tempurl%
\url{https://doi.org/10.1080/14036096.2018.1465994}
\showDOI{\tempurl}


\bibitem[Nethercote(2023)]%
        {nethercote2023platform}
\bibfield{author}{\bibinfo{person}{Megan Nethercote}.} \bibinfo{year}{2023}\natexlab{}.
\newblock \showarticletitle{Platform landlords: Renters, personal data and new digital footholds of urban control}.
\newblock \bibinfo{journal}{\emph{Digital Geography and Society}}  \bibinfo{volume}{5}, Article \bibinfo{articleno}{100060} (\bibinfo{year}{2023}), \bibinfo{numpages}{15}~pages.
\newblock
\urldef\tempurl%
\url{https://doi.org/10.1016/j.diggeo.2023.100060}
\showDOI{\tempurl}


\bibitem[Nissenbaum(2004)]%
        {nissenbaum2004privacy}
\bibfield{author}{\bibinfo{person}{Helen Nissenbaum}.} \bibinfo{year}{2004}\natexlab{}.
\newblock \showarticletitle{Privacy as Contextual Integrity}.
\newblock \bibinfo{journal}{\emph{Washington Law Review}} \bibinfo{volume}{79}, \bibinfo{number}{1} (\bibinfo{year}{2004}), \bibinfo{pages}{119--158}.
\newblock
\urldef\tempurl%
\url{https://digitalcommons.law.uw.edu/wlr/vol79/iss1/10}
\showURL{%
\tempurl}


\bibitem[Parikh and Patel(2017)]%
        {parikh2017cyber}
\bibfield{author}{\bibinfo{person}{Tushar~P Parikh} {and} \bibinfo{person}{Ashok~R Patel}.} \bibinfo{year}{2017}\natexlab{}.
\newblock \showarticletitle{Cyber security: Study on attack, threat, vulnerability}.
\newblock \bibinfo{journal}{\emph{2017 International Journal of Research in Modern Engineering and Emerging Technology}} \bibinfo{volume}{5}, \bibinfo{number}{6} (\bibinfo{year}{2017}).
\newblock


\bibitem[Park et~al\mbox{.}(2018a)]%
        {PARK2018RedditUse}
\bibfield{author}{\bibinfo{person}{Albert Park}, \bibinfo{person}{Mike Conway}, {and} \bibinfo{person}{Annie~T. Chen}.} \bibinfo{year}{2018}\natexlab{a}.
\newblock \showarticletitle{Examining thematic similarity, difference, and membership in three online mental health communities from reddit: A text mining and visualization approach}.
\newblock \bibinfo{journal}{\emph{Computers in Human Behavior}}  \bibinfo{volume}{78} (\bibinfo{year}{2018}), \bibinfo{pages}{98--112}.
\newblock
\urldef\tempurl%
\url{https://doi.org/10.1016/j.chb.2017.09.001}
\showDOI{\tempurl}


\bibitem[Park et~al\mbox{.}(2018b)]%
        {park2018share}
\bibfield{author}{\bibinfo{person}{Cheul~Young Park}, \bibinfo{person}{Cori Faklaris}, \bibinfo{person}{Siyan Zhao}, \bibinfo{person}{Alex Sciuto}, \bibinfo{person}{Laura Dabbish}, {and} \bibinfo{person}{Jason Hong}.} \bibinfo{year}{2018}\natexlab{b}.
\newblock \showarticletitle{Share and Share Alike? An Exploration of Secure Behaviors in Romantic Relationships}. In \bibinfo{booktitle}{\emph{Proceedings of the 14th Symposium on Usable Privacy and Security}}. \bibinfo{publisher}{USENIX Association}, \bibinfo{pages}{83--102}.
\newblock
\urldef\tempurl%
\url{https://www.usenix.org/conference/soups2018/presentation/park}
\showURL{%
\tempurl}


\bibitem[Park and Lim(2020)]%
        {ParkLim_2020_UserExpectations}
\bibfield{author}{\bibinfo{person}{Sunjeong Park} {and} \bibinfo{person}{Youn-kyung Lim}.} \bibinfo{year}{2020}\natexlab{}.
\newblock \showarticletitle{Investigating User Expectations on the Roles of Family-Shared {AI} Speakers}. In \bibinfo{booktitle}{\emph{Proceedings of the 2020 CHI Conference on Human Factors in Computing Systems}}. \bibinfo{publisher}{ACM}, Article \bibinfo{articleno}{323}, \bibinfo{numpages}{13}~pages.
\newblock
\urldef\tempurl%
\url{https://doi.org/10.1145/3313831.3376450}
\showDOI{\tempurl}


\bibitem[Pattnaik et~al\mbox{.}(2023)]%
        {pattnaik2023survey}
\bibfield{author}{\bibinfo{person}{Nandita Pattnaik}, \bibinfo{person}{Shujun Li}, {and} \bibinfo{person}{Jason~R.~C. Nurse}.} \bibinfo{year}{2023}\natexlab{}.
\newblock \showarticletitle{A Survey of User Perspectives on Security and Privacy in a Home Networking Environment}.
\newblock \bibinfo{journal}{\emph{Comput. Surveys}} \bibinfo{volume}{55}, \bibinfo{number}{9}, Article \bibinfo{articleno}{180} (\bibinfo{year}{2023}), \bibinfo{numpages}{38}~pages.
\newblock


\bibitem[Paudel et~al\mbox{.}(2023)]%
        {Riju2023MobilePhone}
\bibfield{author}{\bibinfo{person}{Rizu Paudel}, \bibinfo{person}{Prakriti Dumaru}, \bibinfo{person}{Ankit Shrestha}, \bibinfo{person}{Huzeyfe Kocabas}, {and} \bibinfo{person}{Mahdi~Nasrullah Al-Ameen}.} \bibinfo{year}{2023}\natexlab{}.
\newblock \showarticletitle{A Deep Dive into User's Preferences and Behavior around Mobile Phone Sharing}.
\newblock \bibinfo{journal}{\emph{Proceedings of the ACM on Human-Computer Interaction}} \bibinfo{volume}{7}, \bibinfo{number}{CSCW1}, Article \bibinfo{articleno}{119} (\bibinfo{year}{2023}), \bibinfo{numpages}{22}~pages.
\newblock
\urldef\tempurl%
\url{https://doi.org/10.1145/3579595}
\showDOI{\tempurl}


\bibitem[Proferes et~al\mbox{.}(2021)]%
        {proferes2021SLR-Reddit}
\bibfield{author}{\bibinfo{person}{Nicholas Proferes}, \bibinfo{person}{Naiyan Jones}, \bibinfo{person}{Sarah Gilbert}, \bibinfo{person}{Casey Fiesler}, {and} \bibinfo{person}{Michael Zimmer}.} \bibinfo{year}{2021}\natexlab{}.
\newblock \showarticletitle{Studying {Reddit}: A Systematic Overview of Disciplines, Approaches, Methods, and Ethics}.
\newblock \bibinfo{journal}{\emph{Social Media + Society}} \bibinfo{volume}{7}, \bibinfo{number}{2} (\bibinfo{year}{2021}), \bibinfo{numpages}{14}~pages.
\newblock
\urldef\tempurl%
\url{https://doi.org/10.1177/2056305121101900}
\showDOI{\tempurl}


\bibitem[Reddit(2023)]%
        {Reddit2023Compliance}
\bibfield{author}{\bibinfo{person}{Reddit}.} \bibinfo{year}{2023}\natexlab{}.
\newblock \bibinfo{title}{Data {API} Terms}.
\newblock
\newblock
\urldef\tempurl%
\url{https://www.redditinc.com/policies/data-api-terms}
\showURL{%
\tempurl}


\bibitem[Rocha-Silva et~al\mbox{.}(2023)]%
        {rocha2023passive}
\bibfield{author}{\bibinfo{person}{Tiago Rocha-Silva}, \bibinfo{person}{Concei{\c{c}}{\~a}o Nogueira}, {and} \bibinfo{person}{Liliana Rodrigues}.} \bibinfo{year}{2023}\natexlab{}.
\newblock \showarticletitle{Passive data collection on {Reddit}: a practical approach}.
\newblock \bibinfo{journal}{\emph{Research Ethics}}, Article \bibinfo{articleno}{17470161231210542} (\bibinfo{year}{2023}), \bibinfo{numpages}{18}~pages.
\newblock
\urldef\tempurl%
\url{https://doi.org/10.1177/17470161231210542}
\showDOI{\tempurl}


\bibitem[Ronald et~al\mbox{.}(2023)]%
        {ronald2023institutionalization}
\bibfield{author}{\bibinfo{person}{Richard Ronald}, \bibinfo{person}{Pauline Schijf}, {and} \bibinfo{person}{Kelly Donovan}.} \bibinfo{year}{2023}\natexlab{}.
\newblock \showarticletitle{The institutionalization of shared rental housing and commercial co-living}.
\newblock \bibinfo{journal}{\emph{Housing Studies}} (\bibinfo{year}{2023}), \bibinfo{pages}{1--25}.
\newblock
\urldef\tempurl%
\url{https://doi.org/10.1080/02673037.2023.2176830}
\showDOI{\tempurl}


\bibitem[Sen et~al\mbox{.}(2021)]%
        {sen2021total}
\bibfield{author}{\bibinfo{person}{Indira Sen}, \bibinfo{person}{Fabian Fl{\"o}ck}, \bibinfo{person}{Katrin Weller}, \bibinfo{person}{Bernd Wei{\ss}}, {and} \bibinfo{person}{Claudia Wagner}.} \bibinfo{year}{2021}\natexlab{}.
\newblock \showarticletitle{A total error framework for digital traces of human behavior on online platforms}.
\newblock \bibinfo{journal}{\emph{Public Opinion Quarterly}} \bibinfo{volume}{85}, \bibinfo{number}{S1} (\bibinfo{year}{2021}), \bibinfo{pages}{399--422}.
\newblock


\bibitem[Slupska(2019)]%
        {slupska2019safe}
\bibfield{author}{\bibinfo{person}{Julia Slupska}.} \bibinfo{year}{2019}\natexlab{}.
\newblock \showarticletitle{Safe at home: Towards a feminist critique of cybersecurity}.
\newblock \bibinfo{journal}{\emph{St Antony's International Review}} \bibinfo{volume}{15}, \bibinfo{number}{1} (\bibinfo{year}{2019}), \bibinfo{pages}{83--100}.
\newblock
\urldef\tempurl%
\url{https://www.jstor.org/stable/27027755}
\showURL{%
\tempurl}


\bibitem[Slupska and Tanczer(2021)]%
        {slupska2021threat}
\bibfield{author}{\bibinfo{person}{Julia Slupska} {and} \bibinfo{person}{Leonie~Maria Tanczer}.} \bibinfo{year}{2021}\natexlab{}.
\newblock \showarticletitle{Threat modeling intimate partner violence: tech abuse as a cybersecurity challenge in the {Internet of Things}}.
\newblock In \bibinfo{booktitle}{\emph{The {Emerald} International Handbook of Technology-Facilitated Violence and Abuse}}. \bibinfo{publisher}{Emerald}.
\newblock
\urldef\tempurl%
\url{https://doi.org/10.1108/978-1-83982-848-520211049}
\showDOI{\tempurl}


\bibitem[Spiti et~al\mbox{.}(2022)]%
        {spiti2022social}
\bibfield{author}{\bibinfo{person}{Julia~Muller Spiti}, \bibinfo{person}{Ellen Davies}, \bibinfo{person}{Paul McLiesh}, {and} \bibinfo{person}{Janet Kelly}.} \bibinfo{year}{2022}\natexlab{}.
\newblock \showarticletitle{How social media data are being used to research the experience of mourning: A scoping review}.
\newblock \bibinfo{journal}{\emph{PLOS ONE}} \bibinfo{volume}{17}, \bibinfo{number}{7}, Article \bibinfo{articleno}{e0271034} (\bibinfo{year}{2022}), \bibinfo{numpages}{25}~pages.
\newblock
\urldef\tempurl%
\url{https://doi.org/10.1371/journal.pone.0271034}
\showDOI{\tempurl}


\bibitem[Statista(2021)]%
        {Statista2021AvgNoOfDevice}
\bibfield{author}{\bibinfo{person}{Statista}.} \bibinfo{year}{2021}\natexlab{}.
\newblock \bibinfo{title}{Average number of devices residents have access to in {UK} households in 2020, by device}.
\newblock
\newblock
\urldef\tempurl%
\url{https://www.statista.com/study/41673/connected-devices-in-the-united-kingdom-uk/}
\showURL{%
\tempurl}


\bibitem[Tabassum and Lipford(2023)]%
        {tabassum2023exploring}
\bibfield{author}{\bibinfo{person}{Madiha Tabassum} {and} \bibinfo{person}{Heather Lipford}.} \bibinfo{year}{2023}\natexlab{}.
\newblock \showarticletitle{Exploring privacy implications of awareness and control mechanisms in smart home devices}.
\newblock \bibinfo{journal}{\emph{Proceedings on Privacy Enhancing Technologies}}  \bibinfo{volume}{1} (\bibinfo{year}{2023}), \bibinfo{pages}{571--588}.
\newblock
\urldef\tempurl%
\url{https://doi.org/10.56553/popets-2023-0033}
\showDOI{\tempurl}


\bibitem[Turner et~al\mbox{.}(2022)]%
        {Turner2022ProlificStudy}
\bibfield{author}{\bibinfo{person}{Sarah Turner}, \bibinfo{person}{Nandita Pattnaik}, \bibinfo{person}{Jason~R.C. Nurse}, {and} \bibinfo{person}{Shujun Li}.} \bibinfo{year}{2022}\natexlab{}.
\newblock \showarticletitle{"You Just Assume It Is In There, I Guess": Understanding UK Families' Application and Knowledge of Smart Home Cyber Security}.
\newblock \bibinfo{journal}{\emph{Proceedings of the ACM on Human-Computer Interaction}} \bibinfo{volume}{6}, \bibinfo{number}{CSCW2}, Article \bibinfo{articleno}{269} (\bibinfo{year}{2022}), \bibinfo{numpages}{34}~pages.
\newblock
\urldef\tempurl%
\url{https://doi.org/10.1145/3555159}
\showDOI{\tempurl}


\bibitem[Uyttebrouck et~al\mbox{.}(2020)]%
        {uyttebrouck2020shared}
\bibfield{author}{\bibinfo{person}{Constance Uyttebrouck}, \bibinfo{person}{Ellen Van~Bueren}, {and} \bibinfo{person}{Jacques Teller}.} \bibinfo{year}{2020}\natexlab{}.
\newblock \showarticletitle{Shared housing for students and young professionals: evolution of a market in need of regulation}.
\newblock \bibinfo{journal}{\emph{Journal of Housing and the Built Environment}} \bibinfo{volume}{35}, \bibinfo{number}{4} (\bibinfo{year}{2020}), \bibinfo{pages}{1017--1035}.
\newblock
\urldef\tempurl%
\url{https://doi.org/10.1007/s10901-020-09778-w}
\showDOI{\tempurl}


\bibitem[Wang and Liu(2023)]%
        {Wang2023RedditUse}
\bibfield{author}{\bibinfo{person}{Juite Wang} {and} \bibinfo{person}{Y.-L. Liu}.} \bibinfo{year}{2023}\natexlab{}.
\newblock \showarticletitle{Deep learning-based social media mining for user experience analysis: A case study of smart home products}.
\newblock \bibinfo{journal}{\emph{Technology in Society}}  \bibinfo{volume}{73}, Article \bibinfo{articleno}{102220} (\bibinfo{year}{2023}), \bibinfo{numpages}{18}~pages.
\newblock
\urldef\tempurl%
\url{https://doi.org/10.1016/j.techsoc.2023.102220}
\showDOI{\tempurl}


\bibitem[Watson et~al\mbox{.}(2020)]%
        {Watson2020MU}
\bibfield{author}{\bibinfo{person}{Hue Watson}, \bibinfo{person}{Eyitemi Moju-Igbene}, \bibinfo{person}{Akanksha Kumari}, {and} \bibinfo{person}{Sauvik Das}.} \bibinfo{year}{2020}\natexlab{}.
\newblock \showarticletitle{``We Hold Each Other Accountable''": Unpacking How Social Groups Approach Cybersecurity and Privacy Together}. In \bibinfo{booktitle}{\emph{Proceedings of the 2020 CHI Conference on Human Factors in Computing Systems}}. \bibinfo{publisher}{ACM}, Article \bibinfo{articleno}{478}, \bibinfo{numpages}{12}~pages.
\newblock
\urldef\tempurl%
\url{https://doi.org/10.1145/3313831.3376605}
\showDOI{\tempurl}


\bibitem[Wickramasinghe and Reinhardt(2019)]%
        {wickramasinghe2019survey}
\bibfield{author}{\bibinfo{person}{Chathurangi~Ishara Wickramasinghe} {and} \bibinfo{person}{Delphine Reinhardt}.} \bibinfo{year}{2019}\natexlab{}.
\newblock \showarticletitle{A Survey-Based Exploration of Users' Awareness and Their Willingness to Protect Their Data with Smart Objects}. In \bibinfo{booktitle}{\emph{Privacy and Identity Management. Data for Better Living: AI and Privacy -- 14th IFIP WG 9.2, 9.6/11.7, 11.6/SIG 9.2.2 International Summer School, Windisch, Switzerland, August 19–23, 2019, Revised Selected Papers}}. \bibinfo{publisher}{Springer}, \bibinfo{pages}{427--446}.
\newblock
\urldef\tempurl%
\url{https://doi.org/10.1007/978-3-030-42504-3_27}
\showDOI{\tempurl}


\bibitem[Wilkinson(2014)]%
        {wilkinson2014compare}
\bibfield{author}{\bibinfo{person}{Eleanor Wilkinson}.} \bibinfo{year}{2014}\natexlab{}.
\newblock \showarticletitle{Single people's geographies of home: intimacy and friendship beyond `the family'}.
\newblock \bibinfo{journal}{\emph{Environment and Planning A}} \bibinfo{volume}{46}, \bibinfo{number}{10} (\bibinfo{year}{2014}), \bibinfo{pages}{2452--2468}.
\newblock
\urldef\tempurl%
\url{https://doi.org/10.1068/a130069p}
\showDOI{\tempurl}


\bibitem[Wilkinson and Ortega-Alc{\'a}zar(2019)]%
        {wilkinson2019stranger}
\bibfield{author}{\bibinfo{person}{Eleanor Wilkinson} {and} \bibinfo{person}{Iliana Ortega-Alc{\'a}zar}.} \bibinfo{year}{2019}\natexlab{}.
\newblock \showarticletitle{Stranger danger? The intersectional impacts of shared housing on young people's health \& wellbeing}.
\newblock \bibinfo{journal}{\emph{Health \& Place}}  \bibinfo{volume}{60}, Article \bibinfo{articleno}{102191} (\bibinfo{year}{2019}), \bibinfo{numpages}{7}~pages.
\newblock
\urldef\tempurl%
\url{https://doi.org/10.1016/j.healthplace.2019.102191}
\showDOI{\tempurl}


\bibitem[Williams and Moser(2019)]%
        {williams2019art}
\bibfield{author}{\bibinfo{person}{Michael Williams} {and} \bibinfo{person}{Tami Moser}.} \bibinfo{year}{2019}\natexlab{}.
\newblock \showarticletitle{The art of coding and thematic exploration in qualitative research}.
\newblock \bibinfo{journal}{\emph{International management review}} \bibinfo{volume}{15}, \bibinfo{number}{1} (\bibinfo{year}{2019}), \bibinfo{pages}{45--55}.
\newblock


\bibitem[Windl and Mayer(2022)]%
        {Windl2022MU}
\bibfield{author}{\bibinfo{person}{Maximiliane Windl} {and} \bibinfo{person}{Sven Mayer}.} \bibinfo{year}{2022}\natexlab{}.
\newblock \showarticletitle{The Skewed Privacy Concerns of Bystanders in Smart Environments}.
\newblock \bibinfo{journal}{\emph{Proceedings of the ACM on Human-Computer Interaction}} \bibinfo{volume}{6}, \bibinfo{number}{MHCI}, Article \bibinfo{articleno}{184} (\bibinfo{year}{2022}), \bibinfo{numpages}{21}~pages.
\newblock
\urldef\tempurl%
\url{https://doi.org/10.1145/3546719}
\showDOI{\tempurl}


\bibitem[Wu et~al\mbox{.}(2022)]%
        {wu2022sok}
\bibfield{author}{\bibinfo{person}{Yuxi Wu}, \bibinfo{person}{W.~Keith Edwards}, {and} \bibinfo{person}{Sauvik Das}.} \bibinfo{year}{2022}\natexlab{}.
\newblock \showarticletitle{{SoK}: Social Cybersecurity}. In \bibinfo{booktitle}{\emph{Proceedings of the 2022 IEEE Symposium on Security and Privacy}}. \bibinfo{publisher}{IEEE}, \bibinfo{pages}{1863--1879}.
\newblock
\urldef\tempurl%
\url{https://doi.org/10.1109/SP46214.2022.9833757}
\showDOI{\tempurl}


\bibitem[Yao et~al\mbox{.}(2019)]%
        {yao2019privacy}
\bibfield{author}{\bibinfo{person}{Yaxing Yao}, \bibinfo{person}{Justin~Reed Basdeo}, \bibinfo{person}{Oriana~Rosata Mcdonough}, {and} \bibinfo{person}{Yang Wang}.} \bibinfo{year}{2019}\natexlab{}.
\newblock \showarticletitle{Privacy Perceptions and Designs of Bystanders in Smart Homes}.
\newblock \bibinfo{journal}{\emph{Proceedings of the ACM on Human-Computer Interaction}} \bibinfo{volume}{3}, \bibinfo{number}{CSCW}, Article \bibinfo{articleno}{59} (\bibinfo{year}{2019}), \bibinfo{numpages}{24}~pages.
\newblock
\urldef\tempurl%
\url{https://doi.org/10.1145/3359161}
\showDOI{\tempurl}


\bibitem[Zeng et~al\mbox{.}(2017)]%
        {zeng2017end}
\bibfield{author}{\bibinfo{person}{Eric Zeng}, \bibinfo{person}{Shrirang Mare}, {and} \bibinfo{person}{Franziska Roesner}.} \bibinfo{year}{2017}\natexlab{}.
\newblock \showarticletitle{End User Security and Privacy Concerns with Smart Homes}. In \bibinfo{booktitle}{\emph{Proceedings of the 13th Symposium on Usable Privacy and Security}}. \bibinfo{publisher}{USENIX Association}, \bibinfo{pages}{65--80}.
\newblock
\urldef\tempurl%
\url{https://www.usenix.org/conference/soups2017/technical-sessions/presentation/zeng}
\showURL{%
\tempurl}


\bibitem[Zeng and Roesner(2019)]%
        {zeng2019HomeUser}
\bibfield{author}{\bibinfo{person}{Eric Zeng} {and} \bibinfo{person}{Franziska Roesner}.} \bibinfo{year}{2019}\natexlab{}.
\newblock \showarticletitle{Understanding and Improving Security and Privacy in Multi-User Smart Homes: A Design Exploration and In-Home User Study}. In \bibinfo{booktitle}{\emph{Proceedings of the 28th USENIX Security Symposium}}. \bibinfo{publisher}{USENIX Association}, \bibinfo{pages}{159--176}.
\newblock
\urldef\tempurl%
\url{https://www.usenix.org/conference/usenixsecurity19/presentation/zeng}
\showURL{%
\tempurl}


\bibitem[Zhao(2020)]%
        {zhao2020unraveling}
\bibfield{author}{\bibinfo{person}{Bo Zhao}.} \bibinfo{year}{2020}\natexlab{}.
\newblock \showarticletitle{Unraveling Home Protection in the {IoT} Age: Living, Mixed Reality, and Home 2.0}.
\newblock \bibinfo{journal}{\emph{Science and Technology Law Review}} \bibinfo{volume}{21}, \bibinfo{number}{1} (\bibinfo{year}{2020}), \bibinfo{pages}{43--80}.
\newblock
\urldef\tempurl%
\url{https://doi.org/10.7916/stlr.v21i1.5763}
\showDOI{\tempurl}


\bibitem[Zimmermann et~al\mbox{.}(2019)]%
        {zimmermann2019assessing}
\bibfield{author}{\bibinfo{person}{Verena Zimmermann}, \bibinfo{person}{Paul Gerber}, \bibinfo{person}{Karola Marky}, \bibinfo{person}{Leon Böck}, {and} \bibinfo{person}{Florian Kirchbuchner}.} \bibinfo{year}{2019}\natexlab{}.
\newblock \showarticletitle{Assessing Users' Privacy and Security Concerns of Smart Home Technologies}.
\newblock \bibinfo{journal}{\emph{i-com}} \bibinfo{volume}{18}, \bibinfo{number}{3} (\bibinfo{year}{2019}), \bibinfo{pages}{197--216}.
\newblock
\urldef\tempurl%
\url{https://doi.org/10.1515/icom-2019-0015}
\showDOI{\tempurl}


\end{thebibliography}

\appendix

\includepdf[pages=-]{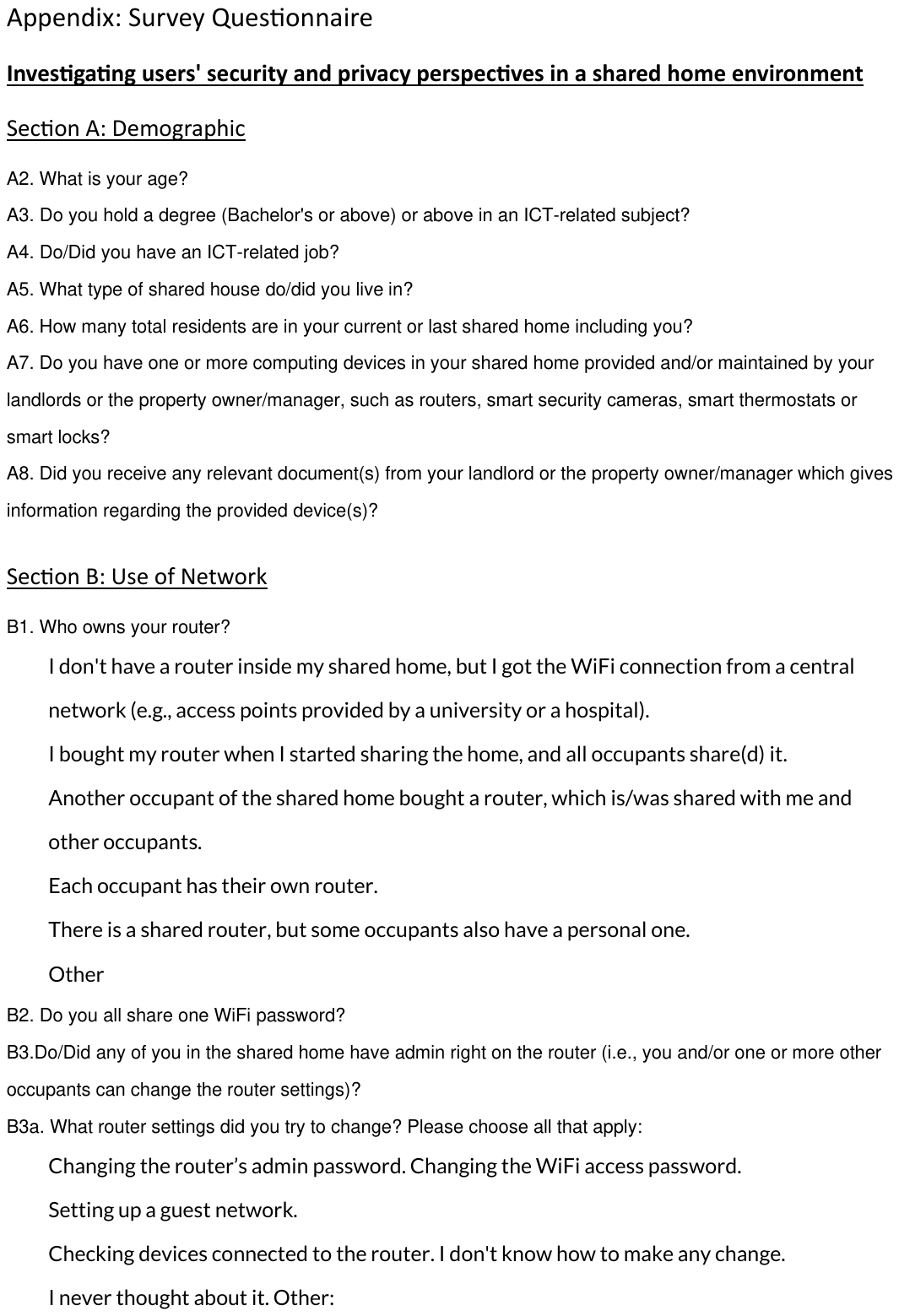}
\includegraphics[scale=.75]{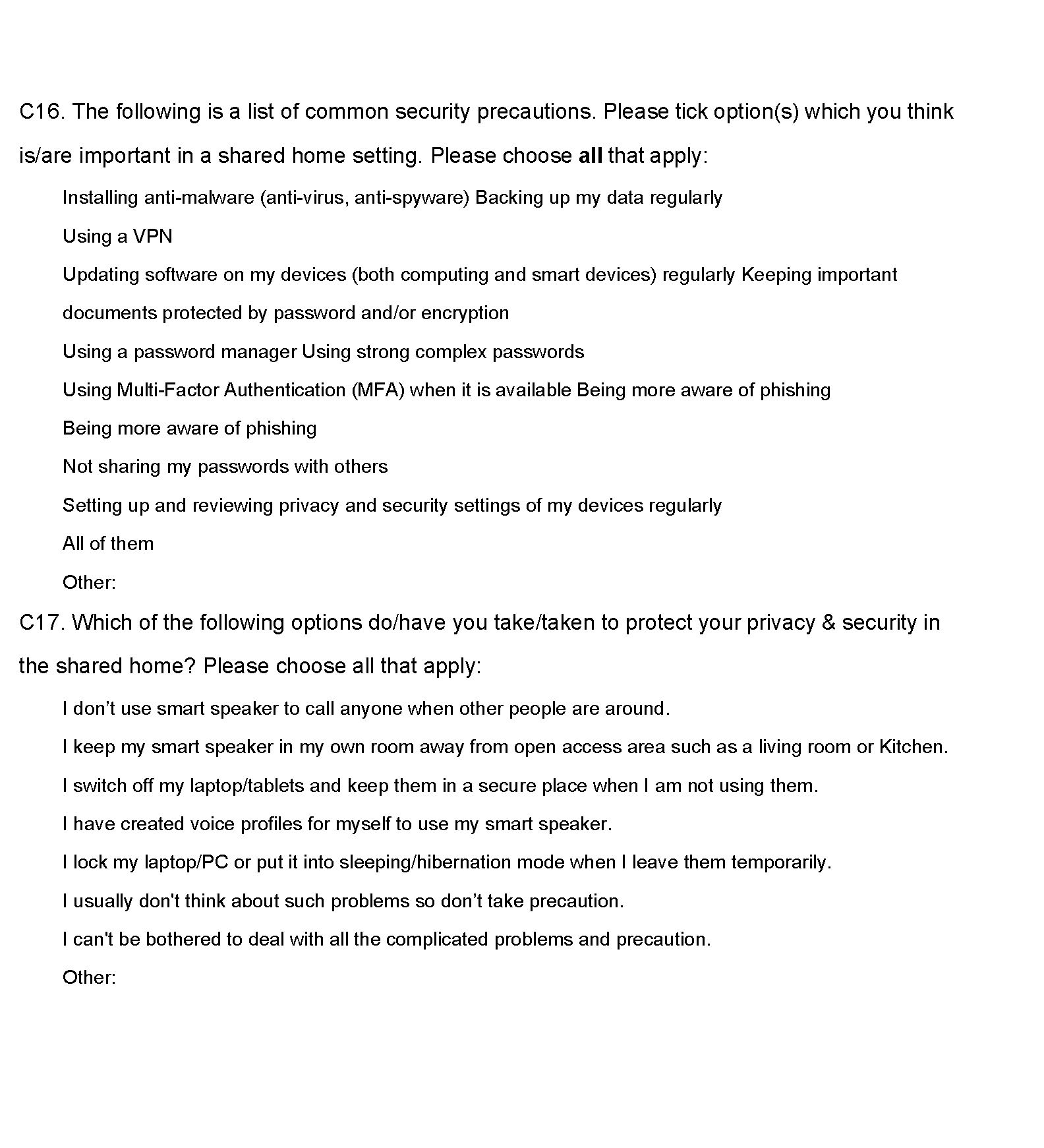}

\nopagebreak

\received{July 2023}
\received[revised]{April 2024}
\received[accepted]{July 2024}

\end{document}